\documentclass[aps,prb,preprint,floatfix,superscriptaddress,showpacs]{revtex4-1}
\usepackage{graphicx}% Include figure files
\usepackage{dcolumn}% Align table columns on decimal point
\usepackage{amsmath,bm}% bold math
\usepackage{color}%
\usepackage{ulem}
\usepackage{multirow}
\usepackage{array} % adjust table width 
\usepackage{xr}

\makeatletter
\newcommand*{\addFileDependency}[1]{% argument=file name and extension
  \typeout{(#1)}
  \@addtofilelist{#1}
  \IfFileExists{#1}{}{\typeout{No file #1.}}
}
\makeatother
 
\newcommand*{\myexternaldocument}[1]{%
    \externaldocument{#1}%
    \addFileDependency{#1.tex}%
    \addFileDependency{#1.aux}%
}

\myexternaldocument{si}

\begin{document}

%\listoftodos
%\newpage

\title{Versatile Physical Properties of a Novel Two-Dimensional Materials Composed of Group IV-V Elements }

\author {Seungjun Lee}
\affiliation{Department of Physics and
             Research Institute for Basic Sciences,
             Kyung Hee University, Seoul, 02447, Korea}

\author{Young-Kyun Kwon}
\email[Corresponding author. E-mail: ]{ykkwon@khu.ac.kr}
\affiliation{Department of Physics and
             Research Institute for Basic Sciences,
             Kyung Hee University, Seoul, 02447, Korea}

\date{\today}
%---------------------------------------------------------------------
\begin{abstract}
Owing to the fascinating physical characteristics of two-dimensional (2D) materials and their heterostructure, much effort has been devoted to exploring their basic physical properties as well as discovering other novel 2D materials.
Herein, based on first-principles calculations, we propose novel 2D material groups with the form A$_2$B$_2$, composed of group IV (A = C, Si, Ge, or Sn) and V (B = N, P, As, Sb, or Bi) elements; the group forms two stable phases with the $P\bar{6}m2$ ($\mathcal{M}$ phase) and $P\bar{3}m1$ ($\mathcal{I}$ phase) crystal symmetries.
We found that a total of 40 different freestanding A$_2$B$_2$ compounds were dynamically stable and displayed versatile physical properties, such as insulating, semiconducting, or metallic properties, depending on their elemental compositions.
Our calculation results further revealed that the newly proposed 2D materials expressed high electrical and thermal transport properties. Additionally, some of the compounds that contained heavy elements exhibited non-trivial topological properties due to strong spin-orbit interactions.
\end{abstract}
%---------------------------------------------------------------------
%Subject Areas: Computational Physics, Condensed Matter Physics, Nanophysics
%\pacs{

\maketitle

%---------------------------------------------------------------------
% If in two-column mode, this environment will change to single-column
% format so that long equations can be displayed. Use sparingly.
%\begin{widetext}
%\end{widetext}

\section{Introduction}
\label{Introduction }
Graphene, the first two-dimensional (2D) material, exhibits fascinating physical properties, such as excellent transport properties,~\cite{{gra1},{gra_thermal}} a very high mechanical strength,~\cite{gra_strength} and a unique electronic structure with massless Dirac fermions.~\cite{{gra2}}
However, since the semi-metallic electronic structure of graphene severely limits its applications, enormous efforts have been devoted to discovering other 2D materials.
As a result, a variety of 2D materials with sizable band gaps have been exfoliated from bulk materials. Examples of these new 2D materials include transition metal dichalcogenides (TMDCs),~\cite{{tmdc1},{tmdc2},{tmdc3},{tmdc4}} phosphorene,~\cite{{toma2014},{Akhtar2017}} and hexagonal boron nitride (h-BN).~\cite{{bn1},{bn2},{bn3}}
%Furthermore, not only experimental efforts but high-throughput computational methods are also actively finding novel 2D materials and discovering their physical properties.~\cite{{Mounet2018},{Haastrup_2018}}
Furthermore, first-principles computational methods have also been successful in finding novel 2D materials and determining their physical properties.~\cite{{Mounet2018},{Haastrup_2018},{chae2019}}
%, due to their high-throughput computational power.

Intriguingly, the electron configurations in 2D semiconductors are distinguishable from bulk semiconductors.
In three-dimensional (3d) semiconductors, atoms usually bond with the four nearest neighbors  in an sp$^3$ tetrahedral structure; this satisfies the octet rule and disallows any dangling bonds or lone pairs of electrons.
As a result, most 3D semiconductors either consist of single elements of group IV, such as Si or Ge, or consist of a combination of group III and V or II and VI elements with 1:1 stoichiometry. 
However, in 2D materials, lone pairs of electrons can be stabilized in the van der Waals gap between atomic layers. 
For example, the P in phosphorene and S in MoS$_2$ have three covalent bonds with the nearest neighbor and one lone pair of electrons pointing out of the atomic layer.
Following the same rule, group III-VI and group IV-V compounds can also form 2D materials with sp$^3$ bonding structures, exhibiting interesting physical properties.
Group III-VI 2D materials in the form of MX (M= Ga or In; X= S, Se, or Te)  have been reported to be semiconductors that exhibit interesting physical properties such as high electron mobility, anomalous optical responses, or topological phase transition by oxygen functionalization.~\cite{{Bandurin2017},{Zhou2018}}
As for the group IV-V compounds, several 2D crystal structures, such as that of the 2D carbon-phosphide group~\cite{{Wang2016},{Guan2016}}, SiP~\cite{SiP}, GeP~\cite{{GeP1},{GeP2}}, or GeAs~\cite{GeAs}, have been theoretically proposed and experimentally synthesized ; however, comprehensive studies of 2D group IV-V compounds have not been conducted.

%Recently, among various 2D materials, layered group IV-V compounds have emerged in both theoretical and experimental studies. For example, it was reported that 2D carbon-phosphide compounds exhibit various physical properties from indirect or direct gap semiconductors, to Dirac semimetal phase depending on their crystal structure.~\cite{{Wang2016},{Guan2016}} Furthermore, various 2D group IV-V compounds such as 2D SiP,~\cite{SiP} GeP~\cite{{GeP1},{GeP2}}, and GeAs~\cite{GeAs}, also have attracted attention to be used for various applications for examples photonic and optoelectronic devices,~\cite{SiP} electronic nanodevices,~\cite{{GeP1},{GeP2}} or photoelectrochemical water splitting appication.~\cite{GeAs}

In this study, we used first-principles density functional theory to propose a novel 2D material group with the form of A$_2$B$_2$, where A and B are elements in group IV (C, Si, Ge, or Sn) and group V (N, P, As, Sb, or Bi), respectively. This new material group contains two distinct phases that possess inversion or mirror symmetry. 
We systematically studied the physical properties, such as geometrical structure, cohesive energy, phonon dispersion relation, and electronic structure of 40 kinds of 2D A$_2$B$_2$ materials.
In addition, to explore the possible areas of application of the selected materials, we calculated their electrical and thermal transport properties and investigated their topological properties.
% 아래 문장 확인 
Intriguingly, we found that 2D A$_2$B$_2$ semiconductors exhibit prominent electron and phonon transport properties, and some of the 2D A$_2$B$_2$ materials containing heavy elements exhibit topological non-trivial properties.
%---------------------------------------------------------------------
\section{Computational details }
\label{Computational}
To investigate the electrical and vibrational properties of various A$_2$B$_2$ structure, we performed first-principles calculations based on density functional theory~\cite{Kohn1965} and density functional perturbation theory (DFPT)~\cite{RevModPhys.73.515} as implemented in the Quantum Espresso (QE) package.~\cite{{QE-2009},{QE-2017}} 
We employed the norm-conserving pseudopotentials~\cite{nc} to describe the valence electrons, and treated exchange-correlation functional within the generalized gradient approximation of  Perdew-Burke-Ernzerhof (PBE).~\cite{Perdew1996} 
The plane-wave kinetic energy cutoff was selected to be 80~Ry, and $c=20$~{\AA} was chosen for the lattice constant of the direction perpendicular to the plane to mimic 2D structure. 
The Brillouin zone (BZ) of each structure was sampled using a $11{\times}11{\times}1$ $k$-point and $6{\times}6{\times}1$ $q$-point grid for the primitive unit cells of A$_2$B$_2$. 
To explore transport properties in A$_2$B$_2$, we solved the  semi-classical Boltzmann transport equation (BTE) for both electrons  and phonons. 
The carrier lifetimes are also calculated by density functional perturbation theory. We calculated anharmonic three phonon (3rd-order) interaction and electron-phonon interaction as scattering sources for phonons and electrons, respectively. 
To account for 3rd-order phonon interaction, we constructed $4{\times}4{\times}1$ supercell and calculated 3rd order inter-atomic force constants using finite-displacement approach. ~\cite{{phonopy},{ShengBTE},{thirdorder}}
For the electron-phonon interaction, we calculated maximum localized  Wannier function~\cite{wannier} and the electron-phonon matrix elements were initially calculated by $10{\times}10{\times}1$ coarse $k$- and $q$-points mesh and were interpolated to $100{\times}100{\times}1$ fine mesh by using EPW package~\cite{EPW}.
Note that the layer thickness of Si$_2$P$_2$ and C$_2$P$_2$, which should be determined to evaluate the conductivity tensor, were choosen 
for 8.28 and 7.39 \AA corresponding to half of out-of-plane lattice constants of their bulk configurations.
%%%5
The topological surface states and $\mathbf{Z_2}$ invariant were calculated by hybrid Wannier charge center method~\cite{wcc}, and maximum localized Wannier function was obtained by WANNIER90 code.~\cite{wannier}

\section{Results and discussion} 
\label{Results}
%\subsection{Crystal and electronic structure}
% entire page. There is no need to do explicit centering.
\begin{figure}[t]
\includegraphics[width=1.0\columnwidth]{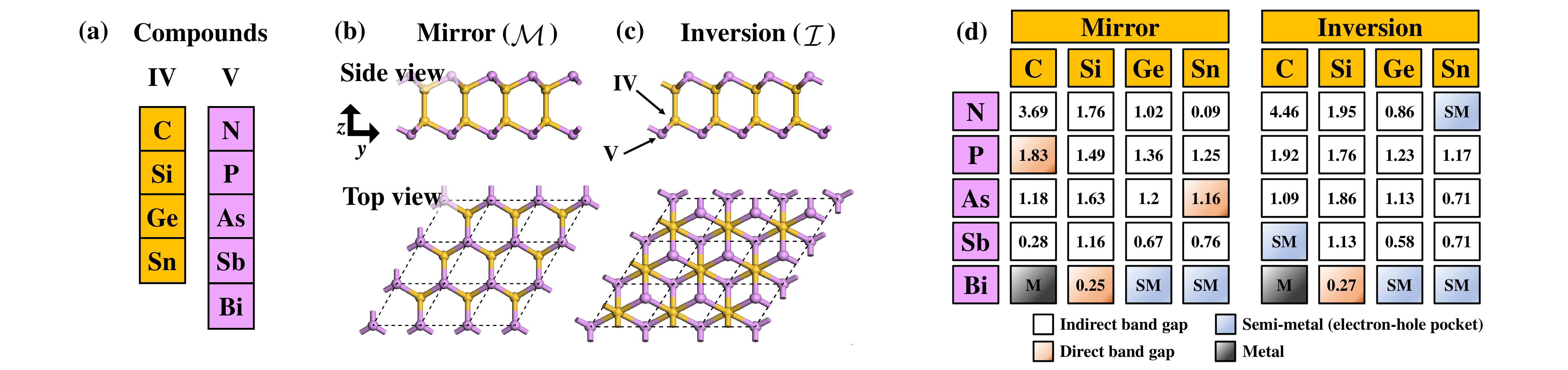}
\caption{(a) Group IV-V elements comprising  the novel 2D materials. Side and top views of the equilibrium structures for two different phases: (b) mirror ($\mathcal{M}$) and (c) inversion ($\mathcal{I}$). The yellow and pink spheres represent the group IV and V elements, respectively, and the dashed line in the top view shows the primitive unit cell of the structure. (d) Electronic structure of A$_2$B$_2$ with the PBE functional. The white and orange boxes represent indirect and direct band gap semiconductors, respectively, with the band gaps measured in terms of electron volts. The sky blue and black boxes indicate semi-metallic and metallic electronic structures, respectively. For, A$_2$Bi$_2$, we considered the spin-orbit coupling (SOC) effect due to strong the SOC strength in Bi; for the other cases, the SOC can be considered to be negligible.
\label{Structure}}
\end{figure}
%---------------------------------------------------------------------

Based on first-principles calculations, we predicted 40 stable compounds formed by the group the IV  element X (X$=$C, Si, Ge, or Sn) and group V element Y (Y$=$N, P, As, Sb, or Bi), with two different phases. 
Figure~\ref{Structure}(a)-(c) show our newly proposed 2D structure (A$_2$B$_2$) and its compound elements. Depending on  the position of the group V elements, A$_2$B$_2$ can form two different phases, one with the mirror symmetry ($\mathcal{M}$ phase) and the other with inversion symmetry ($\mathcal{I}$ phase). 
The crystal symmetries of the $\mathcal{M}$ and the $\mathcal{I}$ phases are $P\bar{6}m2$ and $P\bar{3}m1$, respectively, this relationship is similar to that between the $2H$ and $1T$ phases in TMDCs.~\cite{{tmdc1},{tmdc2}}
We summarize  the calculated equilibrium lattice constants in Table~\ref{table1}.
The $\mathcal{M}$ and $\mathcal{I}$ phases have very similar lattice constants, and the constants are inversely proportional to the atomic numbers.
 
The dynamical stability of all the compounds was confirmed by calculating the phonon dispersion relation.
As shown in figures~\ref{S_phonon1} ($\mathcal{M}$ phase) and~\ref{S_phonon2} ($\mathcal{I}$ phase), there are no imaginary phonon frequencies except in the vicinity  of the $\Gamma$ point.
For carbon-phosphide, it has been reported that the GaSe phase, similar to the $\mathcal{M}$ phase, is the most stable configuration among the various compounds with 1:1 stoichiometry.~\cite{PhysRevB.67.153105} 
Therefore, we confirmed that 2D A$_2$B$_2$ is dynamically as well as energetically stable.
To compare the structural stabilities of the $\mathcal{M}$ and $\mathcal{I}$ phases , we evaluated the cohesive energies of all the structures; they are summarized in Table~\ref{table2}.
We found that in the systems that contain carbon atoms, the $\mathcal{I}$ phase is more stable than the $\mathcal{M}$ phase. In the systems that do not contain carbon atoms, the opposite is true.  
%Furthermore, we compared relative stability of our structure with other recently reported 2D carbon-phosphide system~\ref{{},{}} and  represented their cohesive energies in Talbe S~\ref{supporting materials:cohesive energy}. \subsection{Electronic structure and transport properties} 그냥 언급 안하는게 좋을 듯 

To explore the physical properties of A$_2$B$_2$, we investigated the electronic structures of all the compounds; the results are summarized in Figure~\ref{Structure}(d). 
(Calculated band structures for all A$_2$B$_2$ compounds are presented in Figure~\ref{S_band}.)
We found that most A$_2$B$_2$ compounds are semiconductors, although  some structures exhibit insulating or metallic properties.
Interestingly, the heavier compounds exhibited smaller band gaps.
Furthermore, if the atomic mass of A differed significantly from that of  B,  then A$_2$B$_2$ showed much smaller band gaps; some structures, such as $\mathcal{I}$-Sn$_2$N$_2$ or $\mathcal{I}$-C$_2$Sb$_2$, even became semi-metallic. 
We also note that for the heavier elements, SOC  played an important role in determining the electronic structure. 
The SOC interaction can be considered negligible in A$_2$Sb$_2$ or Sn$_2$B$_2$, but not in A$_2$Bi$_2$. 
The effects of SOC on the A$_2$Bi$_2$ compounds will be discussed in a later section.

% entire page. There is no need to do explicit centering.
\begin{figure}[t]
\includegraphics[width=1.0\columnwidth]{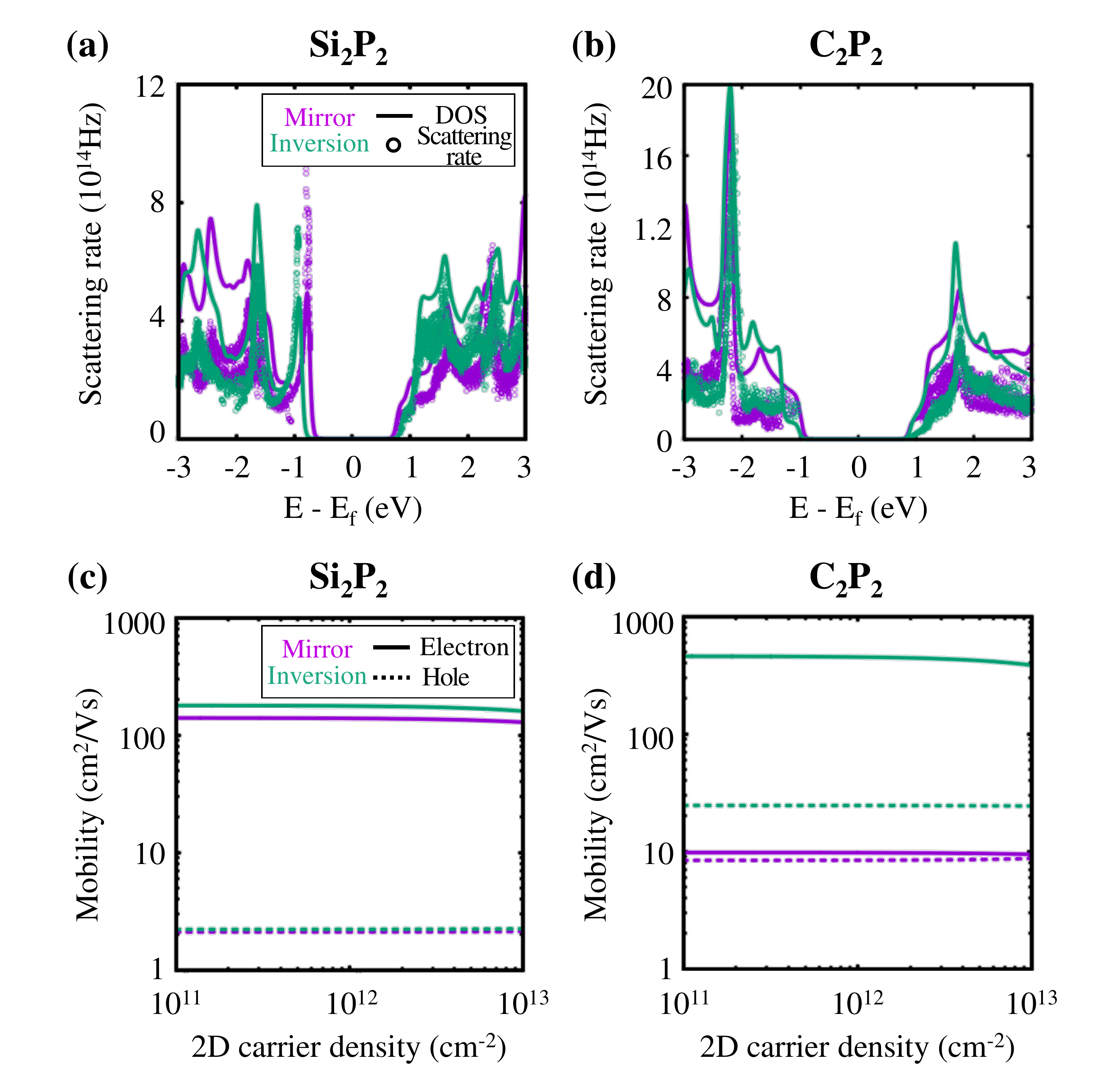}
\caption{Electron-phonon scattering rates of (a) Si$_2$P$_2$ and (b) C$_2$P$_2$ for $\mathcal{M}$ (purple) and $\mathcal{I}$ (cyan) phases with electron densities of states (solid lines). (c) and (d) show the calculated electron (solid line) and hole (dashed line) mobilities as functions of the 2D carrier concentrations of Si$_2$P$_2$ and C$_2$P$_2$, respectively. 
\label{mobility}}
\end{figure}
%---------------------------------------------------------------------

Carrier transport is one of the most important physical properties of semiconductors.
In order to compare the transport properties of A$_2$B$_2$ with those of other 2D semiconductors, we calculated the electrical and thermal conductivities of Si$_2$P$_2$ and C$_2$P$_2$; these  are composed of the elements most common in 2D semiconductors and have band gaps (1.49$\sim$1.92~eV) similar to those of other 2D semiconductors, such as phosphorene (0.88~eV)~\cite{C9CP04372A} and MoS$_2$ (1.9~eV).~\cite{PhysRevB.83.245213}

The carrier mobilities of $\mathcal{M}$-Si$_2$P$_2$ and $\mathcal{M}$-C$_2$P$_2$ have been reported in previous theoretical work by W. Zhang et al~\cite{Zhang2018}. They used the Takagi formula~\cite{takagi} to calculate the carrier mobilities, reporting huge values of room temperature carrier mobilities for $\mathcal{M}$-Si$_2$P$_2$ (46 $\sim$ 8016~cm$^2$s$^{-1}$V$^{-1}$) and $\mathcal{M}$-C$_2$P$_2$ (498 $\sim$ 39289~cm$^2$s$^{-1}$V$^{-1}$).
However, the Takagi formula often significantly overestimates carrier mobilities in 2D materials.~\cite{{Liao2015},{PhysRevB.98.115416},{Slee2020}}
Therefore, to avoid overestimation , we calculated the electron-phonon interaction, which is a main scattering process in the semiconductor, for all band indices ($i$) and crystal momentums ($\mathbf{k}$), as shown in Fig.~\ref{mobility}(a, b). We followed the approach of Liao et al.~\cite{Liao2015}, accounting for the anisotropy of the matrix elements and nonparabolic band structures.
According  to Fermi's golden rule, the electron-phonon scattering rate must be proportional to the available final states, in other words, to the density of states (DOS).
Thus, in both the Si$_2$P$_2$ and C$_2$P$_2$, electrons have longer lifetimes than do holes, due to the lower DOS in the conduction bands.
Based on the calculated carrier lifetimes, we investigated the room-temperature carrier mobilities of Si$_2$P$_2$ and C$_2$P$_2$ according to the carrier types and densities, as shown in Fig.~\ref{mobility}(c, d).
Due to the restriction of electron-phonon scattering in the conduction band, both Si$_2$P$_2$ and C$_2$P$_2$ showed higher carrier mobilities in the $n$ doping region than in the $p$ doping region.
The calculated electron mobilities at a moderate carrier concentration (10$^{12}$~cm$^{-2}$) are 139 and 177~cm$^2$s$^{-1}$V$^{-1}$ for $\mathcal{M}$- and $\mathcal{I}$-Si$_2$P$_2$, and 9.6 and 451~cm$^2$s$^{-1}$V$^{-1}$ for $\mathcal{M}$- and $\mathcal{I}$-C$_2$P$_2$, respectively. These values are comparable to or higher than those of other 2D materials calculated using the same method, such as phosphorene (60$\sim$170~cm$^2$s$^{-1}$V$^{-1}$)~\cite{{Liao2015},{Slee2020}} and MoS$_2$ (26~cm$^2$s$^{-1}$V$^{-1}$).~\cite{MoS2_mu}
Therefore, newly proposed A$_2$B$_2$ compounds can be high performance $n$ type 2D semiconductors, owing to their prominent transport properties.

% entire page. There is no need to do explicit centering.
\begin{figure}[t]
\includegraphics[width=1.0\columnwidth]{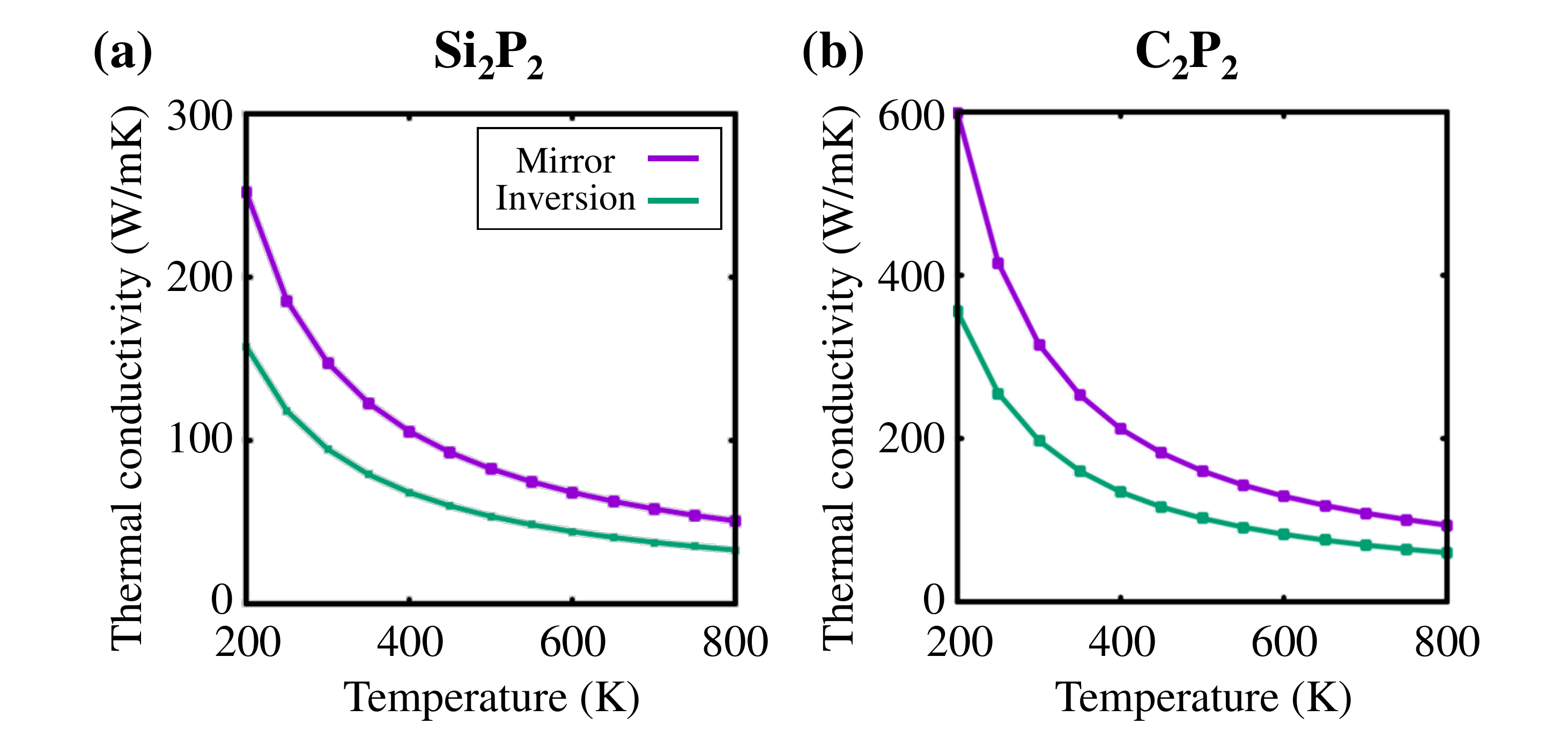}
\caption{Calculated lattice thermal conductivities of (a) Si$_2$P$_2$ and (b) C$_2$P$_2$ as functions of temperature ranging from 200 K to 800 K. The solid purple and cyan lines represent $\mathcal{M}$  and $\mathcal{I}$  phases, respectively. 
\label{kappa}}
\end{figure}
%---------------------------------------------------------------------

The lattice thermal conductivity is also an important property in various applications such as thermoelectric devices.
We calculated the lattice thermal conductivities of Si$_2$P$_2$ and C$_2$P$_2$ by solving the Boltzmann transport equation, as shown in Fig.~\ref{mobility}(e, f). 
We found the room-temperature thermal conductivities to be 146 and 94~Wm$^{-1}$K$^{-1}$ for $\mathcal{M}$- and $\mathcal{I}$-Si$_2$P$_2$, and 313 and 196~Wm$^{-1}$K$^{-1}$ for $\mathcal{M}$- and $\mathcal{I}$-C$_2$P$_2$; these values are slightly larger than those of other 2D materials, such as 
silicene (26~Wm$^{-1}$K$^{-1}$),~\cite{doi:10.1063/1.4905540} 
MoS$_{2}$ (23.2$\sim$34.5~Wm$^{-1}$K$^{-1}$),~\cite{{YC1},{Peng2015},{Yan2014}}
or phosphorene (65$\sim$146~Wm$^{-1}$K$^{-1}$).~\cite{seungjun}
The relatively high thermal conductivities of Si$_2$P$_2$ and C$_2$P$_2$ imply that they cannot be good candidates for thermoelectric materials. 
However, we noticed that the same bonding characteristics of A$_2$B$_2$ may lead to a simple relation between the lattice thermal conductivity and atomic mass; thus, A$_2$B$_2$ covers a wide range of lattice thermal conductivities.
Therefore, due to its versatile physical properties that depend on the compound elements, we believe that A$_2$B$_2$ can be used for various low-dimensional applications.
%Based on our calculation results for Si$_2$P$_2$ and C$_2$P$_2$, we may estimate the physical properties of other A$_2$B$_2$, and also we expected that newly proposed A$_2$B$_2$ can be wide utilized as a high performance $n$ type 2D semiconductors owing to its prominent transport properties.

% entire page. There is no need to do explicit centering.
\begin{figure}[t]
\includegraphics[width=1.0\columnwidth]{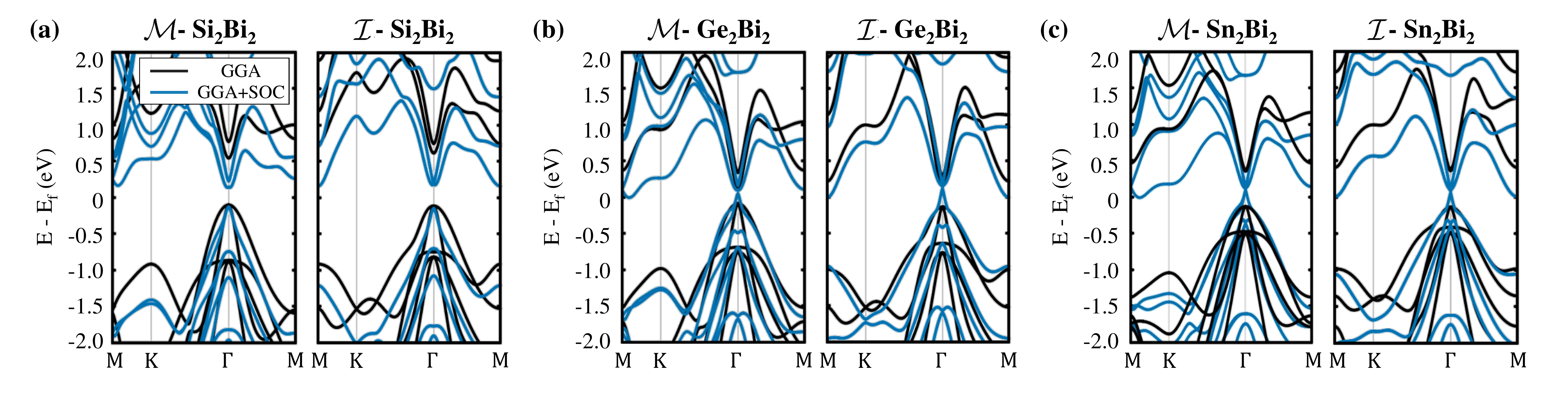}
\caption{Electronic structures of the $\mathcal{M}$ and $\mathcal{I}$ phases of (a) Si$_2$Bi$_2$, (b) Ge$_2$Bi$_2$, and (c) Sn$_2$Bi$_2 $, with (sky blue) and without (black) SOC.
\label{band}}
\end{figure}
%---------------------------------------------------------------------

%In the last section, we investigate the role of SOC interaction 
It has been reported that strong SOC in the low-dimensional system can lead to a topological phase transition via band inversion between the conduction and the valence bands.~\cite{{RevModPhys.83.1057},{RevModPhys.82.3045}}
To further investigate the effects  of SOC in A$_2$B$_2$, we calculated the electronic structures of A$_2$Bi$_2$s (A = Si, Ge, and Sn), which have a strong SOC strength due to Bi.
Figures~\ref{band}(a)-(c) show the electronic structures of A$_2$Bi$_2$ with and without SOC effects.
Without the SOC interaction, A$_2$Bi$_2$s are direct band gap semiconductors with sizable band gap values of 0.3$\sim$0.8~eV.
When the SOC was activated, the calculated band gap was greatly reduced.
For example, the band gap of Si$_2$Bi$_2$ decreased  from 0.8~eV to 0.3~eV.
The band gaps of Ge$_2$Bi$_2$ and Sn$_2$Bi$_2$ were almost closed at $\Gamma$ point, and the compounds exhibited semi-metallic electronic structures. 
Furthermore, the dispersion relation looked almost like linear band dispersion, similar to that of the Dirac cone structure.
%Compared with $\mathcal{M}$ and $\mathcal{I}$ phases, 
%We further noticed that Rashba-type splitting is observed in their band structures regardless of $\mathcal{M}$ and $\mathcal{I}$ phases. 
%it becomes a topological insulator with inverted CB1 and VB1 above a certain higher tensile strain.~\cite{SJL??}

% entire page. There is no need to do explicit centering.
\begin{figure}[t]
\includegraphics[width=1.0\columnwidth]{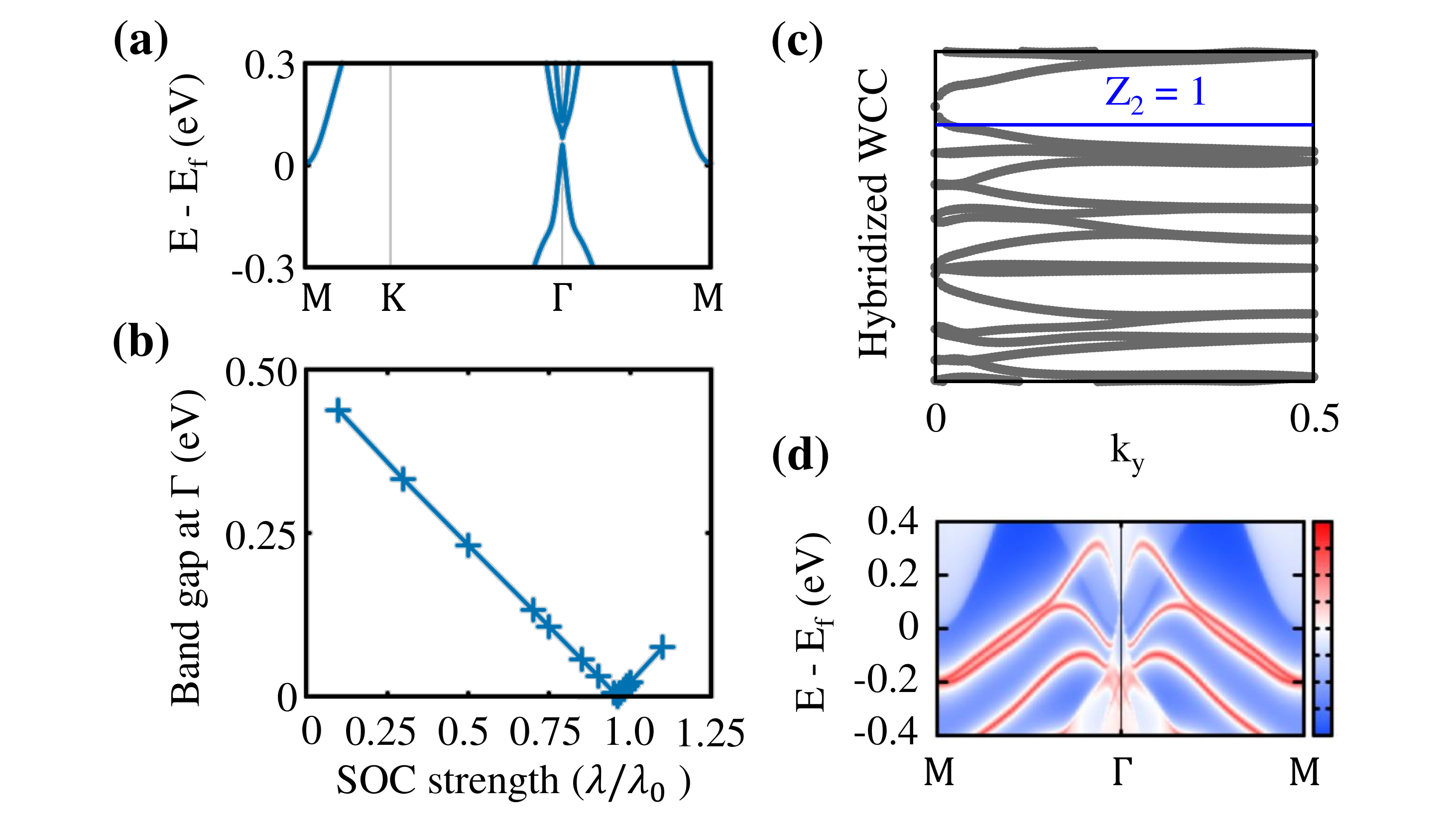}
\caption{(a) A close-up view  of the band structure of $\mathcal{I}$-Sn$_2$Bi$_2$ near the Fermi level, including the SOC
effect, (b) change in the band gap at $\Gamma$ point, according to SOC
strength $\Lambda$, (c) evolution of hybridized Wannier charge center,
and (d) surface band structure of $\mathcal{I}$-Sn$_2$Bi$_2$. 
\label{TI}}
\end{figure}
%---------------------------------------------------------------------

To examine such an interesting dispersion relation in more detail, we scrutinized the band dispersion of $\mathcal{I}$-Sn$_2$Bi$_2$ near the  Fermi level, as shown in Fig.~\ref{TI}(a).
At the $\Gamma$ point, we found a tiny band gap value of 0.02~eV.
To understand the interesting dispersion relation observed in $\mathcal{I}$-Sn$_2$Bi$_2$, we artificially manipulated the SOC strength ($\lambda$) and traced the band gap changes at the $\mathbf{\Gamma}$ point, as represented in Figure~\ref{TI}(b).
We found that the band gap of $\mathcal{I}$-Sn$_2$Bi$_2$ decreases linearly with the increase of the $\lambda$ and closes at $\lambda=0.96\lambda_{0}$; after that point, the valence band and conduction band are reversed and the band gap linearly increases.
It is well known that band inversion with a strong SOC interaction leads to a topological phase transition and the appearance of symmetry protected edge states.~\cite{{RevModPhys.83.1057},{RevModPhys.82.3045}}
To confirm whether or not $\mathcal{I}$-Sn$_2$Bi$_2$ is a topological insulator, we calculated the $\mathbf{Z_2}$ invariant using the hybridized Wannier charge center (HWCC) method.~\cite{wcc} 
Figure~\ref{TI} (c) shows the evolution of the HWCC with increasing $\mathbf{k_y}$. As plotted in Figure~\ref{TI}(c), an arbitrary horizontal line touches the HWCC an odd number of times, which means $\mathbf{Z_2}=1$.  
We also calculated the topological edge states of $\mathcal{I}$-Sn$_2$Bi$_2$ using the surface Green's function method for a semi-infinite system.~\cite{wt} 
Figure~\ref{TI}(d) shows the edge electronic structure of $\mathcal{I}$-Sn$_2$Bi$_2$, where a pair of edge states across the Fermi level moves from the valence band to the conduction band.
Based on these results, we confirmed that $\mathcal{I}$-Sn$_2$Bi$_2$ is a 2D topological insulator.
We also revealed that $\mathcal{M}$-Sn$_2$Bi$_2$ is also a 2D topological insulator (See supplementary Figure~\ref{TI_M}).
As for the other compounds, Ge$_2$Bi$_2$ and Si$_2$Bi$_2$, they are topologically trivial because the SOC is not strong enough to close their relatively large band gaps.
Nevertheless, since strain~\cite{PhysRevB.94.085417} or external fields~\cite{doi:10.1021/nl5043769} reduce the band gaps of 2D materials, Ge$_2$Bi$_2$ and Si$_2$Bi$_2$ may become 2D topological insulators under the correct perturbation .

%Conclusion
\section{Conclusion }
\label{Conclusion}
Based on first-principles density functional theory, we predicted a novel stable 2D material group, A$_2$B$_2$, which is composed of group IV-V elements. 
The newly proposed IV-V compounds have two distinct stable phases--one with mirror symmetry ($\mathcal{M}$-phase), and the other with inversion symmetry ($\mathcal{I}$-phase). 
To explore the physical properties of these compounds, we evaluated the equilibrium structure, cohesive energy, phonon dispersion relations, electronic structure, and electrical and thermal transport properties including the phonon-mediated electron scattering rate and anharmonic phonon interaction.
We found that depending on the combination of elements A and B, 2D A$_2$B$_2$ exhibits versatile physical properties such as large gap insulator, direct or indirect gap semiconductors, topological insulator, or metallic structure.
Our results suggest that due to its fascinating physical properties, such as high-performance electronics and optoelectronics, the newly proposed 2D A$_2$B$_2$ can be used in various applications 

%%Acknowledgements
%\acknowledgments
%This work was supported by ??? grant (??-????-?????????).
%We thank ??? for helpful discussions.

%+++++++++++++++++++++++++++++++++++++++++++++++++++++++++++++++++++++
\bibliographystyle{apsrev}% your bst file here
\bibliography{my.bib} %your bib file here

\begin{thebibliography}{54}
\expandafter\ifx\csname natexlab\endcsname\relax\def\natexlab#1{#1}\fi
\expandafter\ifx\csname bibnamefont\endcsname\relax
  \def\bibnamefont#1{#1}\fi
\expandafter\ifx\csname bibfnamefont\endcsname\relax
  \def\bibfnamefont#1{#1}\fi
\expandafter\ifx\csname citenamefont\endcsname\relax
  \def\citenamefont#1{#1}\fi
\expandafter\ifx\csname url\endcsname\relax
  \def\url#1{\texttt{#1}}\fi
\expandafter\ifx\csname urlprefix\endcsname\relax\def\urlprefix{URL }\fi
\providecommand{\bibinfo}[2]{#2}
\providecommand{\eprint}[2][]{\url{#2}}

\bibitem[{\citenamefont{Novoselov et~al.}(2004)\citenamefont{Novoselov, Geim,
  Morozov, Jiang, Zhang, Dubonos, Grigorieva, and Firsov}}]{gra1}
\bibinfo{author}{\bibfnamefont{K.~S.} \bibnamefont{Novoselov}},
  \bibinfo{author}{\bibfnamefont{A.~K.} \bibnamefont{Geim}},
  \bibinfo{author}{\bibfnamefont{S.~V.} \bibnamefont{Morozov}},
  \bibinfo{author}{\bibfnamefont{D.}~\bibnamefont{Jiang}},
  \bibinfo{author}{\bibfnamefont{Y.}~\bibnamefont{Zhang}},
  \bibinfo{author}{\bibfnamefont{S.~V.} \bibnamefont{Dubonos}},
  \bibinfo{author}{\bibfnamefont{I.~V.} \bibnamefont{Grigorieva}},
  \bibnamefont{and} \bibinfo{author}{\bibfnamefont{A.~A.}
  \bibnamefont{Firsov}}, \bibinfo{journal}{Science}
  \textbf{\bibinfo{volume}{306}}, \bibinfo{pages}{666} (\bibinfo{year}{2004}).

\bibitem[{\citenamefont{Balandin et~al.}(2008)\citenamefont{Balandin, Ghosh,
  Bao, Calizo, Teweldebrhan, Miao, and Lau}}]{gra_thermal}
\bibinfo{author}{\bibfnamefont{A.~A.} \bibnamefont{Balandin}},
  \bibinfo{author}{\bibfnamefont{S.}~\bibnamefont{Ghosh}},
  \bibinfo{author}{\bibfnamefont{W.}~\bibnamefont{Bao}},
  \bibinfo{author}{\bibfnamefont{I.}~\bibnamefont{Calizo}},
  \bibinfo{author}{\bibfnamefont{D.}~\bibnamefont{Teweldebrhan}},
  \bibinfo{author}{\bibfnamefont{F.}~\bibnamefont{Miao}}, \bibnamefont{and}
  \bibinfo{author}{\bibfnamefont{C.~N.} \bibnamefont{Lau}},
  \bibinfo{journal}{Nano Lett.} \textbf{\bibinfo{volume}{8}},
  \bibinfo{pages}{902} (\bibinfo{year}{2008}).

\bibitem[{\citenamefont{Lee et~al.}(2008)\citenamefont{Lee, Wei, Kysar, and
  Hone}}]{gra_strength}
\bibinfo{author}{\bibfnamefont{C.}~\bibnamefont{Lee}},
  \bibinfo{author}{\bibfnamefont{X.}~\bibnamefont{Wei}},
  \bibinfo{author}{\bibfnamefont{J.~W.} \bibnamefont{Kysar}}, \bibnamefont{and}
  \bibinfo{author}{\bibfnamefont{J.}~\bibnamefont{Hone}},
  \bibinfo{journal}{Science} \textbf{\bibinfo{volume}{321}},
  \bibinfo{pages}{385} (\bibinfo{year}{2008}), ISSN \bibinfo{issn}{0036-8075}.

\bibitem[{\citenamefont{Geim and Novoselov}(2007)}]{gra2}
\bibinfo{author}{\bibfnamefont{A.~K.} \bibnamefont{Geim}} \bibnamefont{and}
  \bibinfo{author}{\bibfnamefont{K.~S.} \bibnamefont{Novoselov}},
  \bibinfo{journal}{Nat. Mater.} \textbf{\bibinfo{volume}{6}},
  \bibinfo{pages}{183} (\bibinfo{year}{2007}).

\bibitem[{\citenamefont{Tan and Zhang}(2015)}]{tmdc1}
\bibinfo{author}{\bibfnamefont{C.~L.} \bibnamefont{Tan}} \bibnamefont{and}
  \bibinfo{author}{\bibfnamefont{H.}~\bibnamefont{Zhang}},
  \bibinfo{journal}{Chem. Soc. Rev.} \textbf{\bibinfo{volume}{44}},
  \bibinfo{pages}{2713} (\bibinfo{year}{2015}).

\bibitem[{\citenamefont{Chhowalla et~al.}(2013)\citenamefont{Chhowalla, Shin,
  Eda, Li, Loh, and Zhang}}]{tmdc2}
\bibinfo{author}{\bibfnamefont{M.}~\bibnamefont{Chhowalla}},
  \bibinfo{author}{\bibfnamefont{H.~S.} \bibnamefont{Shin}},
  \bibinfo{author}{\bibfnamefont{G.}~\bibnamefont{Eda}},
  \bibinfo{author}{\bibfnamefont{L.~J.} \bibnamefont{Li}},
  \bibinfo{author}{\bibfnamefont{K.}~\bibnamefont{Loh}}, \bibnamefont{and}
  \bibinfo{author}{\bibfnamefont{H.}~\bibnamefont{Zhang}},
  \bibinfo{journal}{Nat. Chem.} \textbf{\bibinfo{volume}{5}},
  \bibinfo{pages}{263} (\bibinfo{year}{2013}).

\bibitem[{\citenamefont{Huang et~al.}(2013)\citenamefont{Huang, Zeng, and
  Zhang}}]{tmdc3}
\bibinfo{author}{\bibfnamefont{X.}~\bibnamefont{Huang}},
  \bibinfo{author}{\bibfnamefont{Z.~Y.} \bibnamefont{Zeng}}, \bibnamefont{and}
  \bibinfo{author}{\bibfnamefont{H.}~\bibnamefont{Zhang}},
  \bibinfo{journal}{Chem. Soc. Rev.} \textbf{\bibinfo{volume}{42}},
  \bibinfo{pages}{1934} (\bibinfo{year}{2013}).

\bibitem[{\citenamefont{Lv et~al.}(2015)\citenamefont{Lv, Robinson, Schaak,
  Sun, Sun, Mallouk, and Terrones}}]{tmdc4}
\bibinfo{author}{\bibfnamefont{R.}~\bibnamefont{Lv}},
  \bibinfo{author}{\bibfnamefont{J.~A.} \bibnamefont{Robinson}},
  \bibinfo{author}{\bibfnamefont{R.~E.} \bibnamefont{Schaak}},
  \bibinfo{author}{\bibfnamefont{D.}~\bibnamefont{Sun}},
  \bibinfo{author}{\bibfnamefont{Y.}~\bibnamefont{Sun}},
  \bibinfo{author}{\bibfnamefont{T.~E.} \bibnamefont{Mallouk}},
  \bibnamefont{and} \bibinfo{author}{\bibfnamefont{M.}~\bibnamefont{Terrones}},
  \bibinfo{journal}{Acc. Chem. Res.} \textbf{\bibinfo{volume}{48}},
  \bibinfo{pages}{56} (\bibinfo{year}{2015}).

\bibitem[{\citenamefont{Liu et~al.}(2014)\citenamefont{Liu, Neal, Zhu, Luo, Xu,
  Tom\'anek, and Ye}}]{toma2014}
\bibinfo{author}{\bibfnamefont{H.}~\bibnamefont{Liu}},
  \bibinfo{author}{\bibfnamefont{A.~T.} \bibnamefont{Neal}},
  \bibinfo{author}{\bibfnamefont{Z.}~\bibnamefont{Zhu}},
  \bibinfo{author}{\bibfnamefont{Z.}~\bibnamefont{Luo}},
  \bibinfo{author}{\bibfnamefont{X.}~\bibnamefont{Xu}},
  \bibinfo{author}{\bibfnamefont{D.}~\bibnamefont{Tom\'anek}},
  \bibnamefont{and} \bibinfo{author}{\bibfnamefont{P.~D.} \bibnamefont{Ye}},
  \bibinfo{journal}{ACS Nano} \textbf{\bibinfo{volume}{8}},
  \bibinfo{pages}{4033} (\bibinfo{year}{2014}).

\bibitem[{\citenamefont{Akhtar et~al.}(2017)\citenamefont{Akhtar, Anderson,
  Zhao, Alruqi, Mroczkowska, Sumanasekera, and Jasinski}}]{Akhtar2017}
\bibinfo{author}{\bibfnamefont{M.}~\bibnamefont{Akhtar}},
  \bibinfo{author}{\bibfnamefont{G.}~\bibnamefont{Anderson}},
  \bibinfo{author}{\bibfnamefont{R.}~\bibnamefont{Zhao}},
  \bibinfo{author}{\bibfnamefont{A.}~\bibnamefont{Alruqi}},
  \bibinfo{author}{\bibfnamefont{J.~E.} \bibnamefont{Mroczkowska}},
  \bibinfo{author}{\bibfnamefont{G.}~\bibnamefont{Sumanasekera}},
  \bibnamefont{and} \bibinfo{author}{\bibfnamefont{J.~B.}
  \bibnamefont{Jasinski}}, \bibinfo{journal}{npj 2D Mater. Appl.}
  \textbf{\bibinfo{volume}{1}}, \bibinfo{pages}{5} (\bibinfo{year}{2017}).

\bibitem[{\citenamefont{Lin et~al.}(2010)\citenamefont{Lin, Williams, and
  Connell}}]{bn1}
\bibinfo{author}{\bibfnamefont{Y.}~\bibnamefont{Lin}},
  \bibinfo{author}{\bibfnamefont{T.~V.} \bibnamefont{Williams}},
  \bibnamefont{and} \bibinfo{author}{\bibfnamefont{J.~W.~S.}
  \bibnamefont{Connell}}, \bibinfo{journal}{J. Phys. Chem. Lett.}
  \textbf{\bibinfo{volume}{1}}, \bibinfo{pages}{277} (\bibinfo{year}{2010}).

\bibitem[{\citenamefont{Weng et~al.}(2016)\citenamefont{Weng, Wang, Wang,
  Bando, and Golberg}}]{bn2}
\bibinfo{author}{\bibfnamefont{Q.}~\bibnamefont{Weng}},
  \bibinfo{author}{\bibfnamefont{X.}~\bibnamefont{Wang}},
  \bibinfo{author}{\bibfnamefont{X.}~\bibnamefont{Wang}},
  \bibinfo{author}{\bibfnamefont{Y.}~\bibnamefont{Bando}}, \bibnamefont{and}
  \bibinfo{author}{\bibfnamefont{D.}~\bibnamefont{Golberg}},
  \bibinfo{journal}{Chem. Soc. Rev.} \textbf{\bibinfo{volume}{45}},
  \bibinfo{pages}{3989} (\bibinfo{year}{2016}).

\bibitem[{\citenamefont{Li and Chen}(2016)}]{bn3}
\bibinfo{author}{\bibfnamefont{L.~H.} \bibnamefont{Li}} \bibnamefont{and}
  \bibinfo{author}{\bibfnamefont{Y.}~\bibnamefont{Chen}},
  \bibinfo{journal}{Adv. Funct. Mater.} \textbf{\bibinfo{volume}{26}},
  \bibinfo{pages}{2594} (\bibinfo{year}{2016}).

\bibitem[{\citenamefont{Mounet et~al.}(2018)\citenamefont{Mounet, Gibertini,
  Schwaller, Campi, Merkys, Marrazzo, Sohier, Castelli, Cepellotti, Pizzi
  et~al.}}]{Mounet2018}
\bibinfo{author}{\bibfnamefont{N.}~\bibnamefont{Mounet}},
  \bibinfo{author}{\bibfnamefont{M.}~\bibnamefont{Gibertini}},
  \bibinfo{author}{\bibfnamefont{P.}~\bibnamefont{Schwaller}},
  \bibinfo{author}{\bibfnamefont{D.}~\bibnamefont{Campi}},
  \bibinfo{author}{\bibfnamefont{A.}~\bibnamefont{Merkys}},
  \bibinfo{author}{\bibfnamefont{A.}~\bibnamefont{Marrazzo}},
  \bibinfo{author}{\bibfnamefont{T.}~\bibnamefont{Sohier}},
  \bibinfo{author}{\bibfnamefont{I.~E.} \bibnamefont{Castelli}},
  \bibinfo{author}{\bibfnamefont{A.}~\bibnamefont{Cepellotti}},
  \bibinfo{author}{\bibfnamefont{G.}~\bibnamefont{Pizzi}},
  \bibnamefont{et~al.}, \bibinfo{journal}{Nat. Nanotechnol.}
  \textbf{\bibinfo{volume}{13}}, \bibinfo{pages}{246} (\bibinfo{year}{2018}),
  ISSN \bibinfo{issn}{1748-3395}.

\bibitem[{\citenamefont{Haastrup et~al.}(2018)\citenamefont{Haastrup, Strange,
  Pandey, Deilmann, Schmidt, Hinsche, Gjerding, Torelli, Larsen, Riis-Jensen
  et~al.}}]{Haastrup_2018}
\bibinfo{author}{\bibfnamefont{S.}~\bibnamefont{Haastrup}},
  \bibinfo{author}{\bibfnamefont{M.}~\bibnamefont{Strange}},
  \bibinfo{author}{\bibfnamefont{M.}~\bibnamefont{Pandey}},
  \bibinfo{author}{\bibfnamefont{T.}~\bibnamefont{Deilmann}},
  \bibinfo{author}{\bibfnamefont{P.~S.} \bibnamefont{Schmidt}},
  \bibinfo{author}{\bibfnamefont{N.~F.} \bibnamefont{Hinsche}},
  \bibinfo{author}{\bibfnamefont{M.~N.} \bibnamefont{Gjerding}},
  \bibinfo{author}{\bibfnamefont{D.}~\bibnamefont{Torelli}},
  \bibinfo{author}{\bibfnamefont{P.~M.} \bibnamefont{Larsen}},
  \bibinfo{author}{\bibfnamefont{A.~C.} \bibnamefont{Riis-Jensen}},
  \bibnamefont{et~al.}, \bibinfo{journal}{2D Mater.}
  \textbf{\bibinfo{volume}{5}}, \bibinfo{pages}{042002} (\bibinfo{year}{2018}).

\bibitem[{\citenamefont{Chae and Son}(2019)}]{chae2019}
\bibinfo{author}{\bibfnamefont{K.}~\bibnamefont{Chae}} \bibnamefont{and}
  \bibinfo{author}{\bibfnamefont{Y.-W.} \bibnamefont{Son}},
  \bibinfo{journal}{Nano Lett.} \textbf{\bibinfo{volume}{19}},
  \bibinfo{pages}{2694} (\bibinfo{year}{2019}), ISSN \bibinfo{issn}{1530-6984}.

\bibitem[{\citenamefont{Bandurin et~al.}(2017)\citenamefont{Bandurin, Tyurnina,
  Yu, Mishchenko, Z{\'o}lyomi, Morozov, Kumar, Gorbachev, Kudrynskyi, Pezzini
  et~al.}}]{Bandurin2017}
\bibinfo{author}{\bibfnamefont{D.~A.} \bibnamefont{Bandurin}},
  \bibinfo{author}{\bibfnamefont{A.~V.} \bibnamefont{Tyurnina}},
  \bibinfo{author}{\bibfnamefont{G.~L.} \bibnamefont{Yu}},
  \bibinfo{author}{\bibfnamefont{A.}~\bibnamefont{Mishchenko}},
  \bibinfo{author}{\bibfnamefont{V.}~\bibnamefont{Z{\'o}lyomi}},
  \bibinfo{author}{\bibfnamefont{S.~V.} \bibnamefont{Morozov}},
  \bibinfo{author}{\bibfnamefont{R.~K.} \bibnamefont{Kumar}},
  \bibinfo{author}{\bibfnamefont{R.~V.} \bibnamefont{Gorbachev}},
  \bibinfo{author}{\bibfnamefont{Z.~R.} \bibnamefont{Kudrynskyi}},
  \bibinfo{author}{\bibfnamefont{S.}~\bibnamefont{Pezzini}},
  \bibnamefont{et~al.}, \bibinfo{journal}{Nature Nanotechnol.}
  \textbf{\bibinfo{volume}{12}}, \bibinfo{pages}{223} (\bibinfo{year}{2017}),
  ISSN \bibinfo{issn}{1748-3395}.

\bibitem[{\citenamefont{Zhou et~al.}(2018)\citenamefont{Zhou, Liu, Zhao, and
  Yao}}]{Zhou2018}
\bibinfo{author}{\bibfnamefont{S.}~\bibnamefont{Zhou}},
  \bibinfo{author}{\bibfnamefont{C.-C.} \bibnamefont{Liu}},
  \bibinfo{author}{\bibfnamefont{J.}~\bibnamefont{Zhao}}, \bibnamefont{and}
  \bibinfo{author}{\bibfnamefont{Y.}~\bibnamefont{Yao}}, \bibinfo{journal}{npj
  Quantum Mater.} \textbf{\bibinfo{volume}{3}}, \bibinfo{pages}{16}
  (\bibinfo{year}{2018}), ISSN \bibinfo{issn}{2397-4648}.

\bibitem[{\citenamefont{Wang et~al.}(2016)\citenamefont{Wang, Pandey, and
  Karna}}]{Wang2016}
\bibinfo{author}{\bibfnamefont{G.}~\bibnamefont{Wang}},
  \bibinfo{author}{\bibfnamefont{R.}~\bibnamefont{Pandey}}, \bibnamefont{and}
  \bibinfo{author}{\bibfnamefont{S.~P.} \bibnamefont{Karna}},
  \bibinfo{journal}{Nanoscale} \textbf{\bibinfo{volume}{8}},
  \bibinfo{pages}{8819} (\bibinfo{year}{2016}), ISSN \bibinfo{issn}{20403372}.

\bibitem[{\citenamefont{Guan et~al.}(2016)\citenamefont{Guan, Liu, Zhu, and
  Tom{\'{a}}nek}}]{Guan2016}
\bibinfo{author}{\bibfnamefont{J.}~\bibnamefont{Guan}},
  \bibinfo{author}{\bibfnamefont{D.}~\bibnamefont{Liu}},
  \bibinfo{author}{\bibfnamefont{Z.}~\bibnamefont{Zhu}}, \bibnamefont{and}
  \bibinfo{author}{\bibfnamefont{D.}~\bibnamefont{Tom{\'{a}}nek}},
  \bibinfo{journal}{Nano Letters} \textbf{\bibinfo{volume}{16}},
  \bibinfo{pages}{3247} (\bibinfo{year}{2016}), ISSN \bibinfo{issn}{15306992}.

\bibitem[{\citenamefont{Li et~al.}(2018{\natexlab{a}})\citenamefont{Li, Wang,
  Li, Yu, Jia, Qiao, Zhu, Liu, and Tao}}]{SiP}
\bibinfo{author}{\bibfnamefont{C.}~\bibnamefont{Li}},
  \bibinfo{author}{\bibfnamefont{S.}~\bibnamefont{Wang}},
  \bibinfo{author}{\bibfnamefont{C.}~\bibnamefont{Li}},
  \bibinfo{author}{\bibfnamefont{T.}~\bibnamefont{Yu}},
  \bibinfo{author}{\bibfnamefont{N.}~\bibnamefont{Jia}},
  \bibinfo{author}{\bibfnamefont{J.}~\bibnamefont{Qiao}},
  \bibinfo{author}{\bibfnamefont{M.}~\bibnamefont{Zhu}},
  \bibinfo{author}{\bibfnamefont{D.}~\bibnamefont{Liu}}, \bibnamefont{and}
  \bibinfo{author}{\bibfnamefont{X.}~\bibnamefont{Tao}}, \bibinfo{journal}{J.
  Mater. Chem. C} \textbf{\bibinfo{volume}{6}}, \bibinfo{pages}{7219}
  (\bibinfo{year}{2018}{\natexlab{a}}).

\bibitem[{\citenamefont{Li et~al.}(2018{\natexlab{b}})\citenamefont{Li, Wang,
  Gong, Zhu, Deng, Shi, Gao, Li, and Zhai}}]{GeP1}
\bibinfo{author}{\bibfnamefont{L.}~\bibnamefont{Li}},
  \bibinfo{author}{\bibfnamefont{W.}~\bibnamefont{Wang}},
  \bibinfo{author}{\bibfnamefont{P.}~\bibnamefont{Gong}},
  \bibinfo{author}{\bibfnamefont{X.}~\bibnamefont{Zhu}},
  \bibinfo{author}{\bibfnamefont{B.}~\bibnamefont{Deng}},
  \bibinfo{author}{\bibfnamefont{X.}~\bibnamefont{Shi}},
  \bibinfo{author}{\bibfnamefont{G.}~\bibnamefont{Gao}},
  \bibinfo{author}{\bibfnamefont{H.}~\bibnamefont{Li}}, \bibnamefont{and}
  \bibinfo{author}{\bibfnamefont{T.}~\bibnamefont{Zhai}},
  \bibinfo{journal}{Adv. Mater.} \textbf{\bibinfo{volume}{30}},
  \bibinfo{pages}{1706771} (\bibinfo{year}{2018}{\natexlab{b}}).

\bibitem[{\citenamefont{Li et~al.}(2019)\citenamefont{Li, Shi, He, Ouyang, Li,
  Zhang, Zhang, Tang, Römer, and Zhong}}]{GeP2}
\bibinfo{author}{\bibfnamefont{Z.}~\bibnamefont{Li}},
  \bibinfo{author}{\bibfnamefont{X.}~\bibnamefont{Shi}},
  \bibinfo{author}{\bibfnamefont{C.}~\bibnamefont{He}},
  \bibinfo{author}{\bibfnamefont{T.}~\bibnamefont{Ouyang}},
  \bibinfo{author}{\bibfnamefont{J.}~\bibnamefont{Li}},
  \bibinfo{author}{\bibfnamefont{C.}~\bibnamefont{Zhang}},
  \bibinfo{author}{\bibfnamefont{S.}~\bibnamefont{Zhang}},
  \bibinfo{author}{\bibfnamefont{C.}~\bibnamefont{Tang}},
  \bibinfo{author}{\bibfnamefont{R.~A.} \bibnamefont{Römer}},
  \bibnamefont{and} \bibinfo{author}{\bibfnamefont{J.}~\bibnamefont{Zhong}},
  \bibinfo{journal}{Appl. Surf. Sci.} \textbf{\bibinfo{volume}{497}},
  \bibinfo{pages}{143803} (\bibinfo{year}{2019}), ISSN
  \bibinfo{issn}{0169-4332}.

\bibitem[{\citenamefont{Jung et~al.}(2018)\citenamefont{Jung, Kim, Cha, Myung,
  Shojaei, Abbas, Lee, Cha, Park, and Kang}}]{GeAs}
\bibinfo{author}{\bibfnamefont{C.~S.} \bibnamefont{Jung}},
  \bibinfo{author}{\bibfnamefont{D.}~\bibnamefont{Kim}},
  \bibinfo{author}{\bibfnamefont{S.}~\bibnamefont{Cha}},
  \bibinfo{author}{\bibfnamefont{Y.}~\bibnamefont{Myung}},
  \bibinfo{author}{\bibfnamefont{F.}~\bibnamefont{Shojaei}},
  \bibinfo{author}{\bibfnamefont{H.~G.} \bibnamefont{Abbas}},
  \bibinfo{author}{\bibfnamefont{J.~A.} \bibnamefont{Lee}},
  \bibinfo{author}{\bibfnamefont{E.~H.} \bibnamefont{Cha}},
  \bibinfo{author}{\bibfnamefont{J.}~\bibnamefont{Park}}, \bibnamefont{and}
  \bibinfo{author}{\bibfnamefont{H.~S.} \bibnamefont{Kang}},
  \bibinfo{journal}{J. Mater. Chem. A} \textbf{\bibinfo{volume}{6}},
  \bibinfo{pages}{9089} (\bibinfo{year}{2018}).

\bibitem[{\citenamefont{Kohn and Sham}(1965)}]{Kohn1965}
\bibinfo{author}{\bibfnamefont{W.}~\bibnamefont{Kohn}} \bibnamefont{and}
  \bibinfo{author}{\bibfnamefont{L.~J.} \bibnamefont{Sham}},
  \bibinfo{journal}{Phys. Rev.} \textbf{\bibinfo{volume}{140}},
  \bibinfo{pages}{A1133} (\bibinfo{year}{1965}).

\bibitem[{\citenamefont{Baroni et~al.}(2001)\citenamefont{Baroni, de~Gironcoli,
  Dal~Corso, and Giannozzi}}]{RevModPhys.73.515}
\bibinfo{author}{\bibfnamefont{S.}~\bibnamefont{Baroni}},
  \bibinfo{author}{\bibfnamefont{S.}~\bibnamefont{de~Gironcoli}},
  \bibinfo{author}{\bibfnamefont{A.}~\bibnamefont{Dal~Corso}},
  \bibnamefont{and}
  \bibinfo{author}{\bibfnamefont{P.}~\bibnamefont{Giannozzi}},
  \bibinfo{journal}{Rev. Mod. Phys.} \textbf{\bibinfo{volume}{73}},
  \bibinfo{pages}{515} (\bibinfo{year}{2001}).

\bibitem[{\citenamefont{Giannozzi et~al.}(2009)\citenamefont{Giannozzi, Baroni,
  Bonini, Calandra, Car, Cavazzoni, Ceresoli, Chiarotti, Cococcioni, Dabo
  et~al.}}]{QE-2009}
\bibinfo{author}{\bibfnamefont{P.}~\bibnamefont{Giannozzi}},
  \bibinfo{author}{\bibfnamefont{S.}~\bibnamefont{Baroni}},
  \bibinfo{author}{\bibfnamefont{N.}~\bibnamefont{Bonini}},
  \bibinfo{author}{\bibfnamefont{M.}~\bibnamefont{Calandra}},
  \bibinfo{author}{\bibfnamefont{R.}~\bibnamefont{Car}},
  \bibinfo{author}{\bibfnamefont{C.}~\bibnamefont{Cavazzoni}},
  \bibinfo{author}{\bibfnamefont{D.}~\bibnamefont{Ceresoli}},
  \bibinfo{author}{\bibfnamefont{G.~L.} \bibnamefont{Chiarotti}},
  \bibinfo{author}{\bibfnamefont{M.}~\bibnamefont{Cococcioni}},
  \bibinfo{author}{\bibfnamefont{I.}~\bibnamefont{Dabo}}, \bibnamefont{et~al.},
  \bibinfo{journal}{J. Phys. Condens. Matter} \textbf{\bibinfo{volume}{21}},
  \bibinfo{pages}{395502 (19pp)} (\bibinfo{year}{2009}).

\bibitem[{\citenamefont{Giannozzi et~al.}(2017)\citenamefont{Giannozzi,
  Andreussi, Brumme, Bunau, Nardelli, Calandra, Car, Cavazzoni, Ceresoli,
  Cococcioni et~al.}}]{QE-2017}
\bibinfo{author}{\bibfnamefont{P.}~\bibnamefont{Giannozzi}},
  \bibinfo{author}{\bibfnamefont{O.}~\bibnamefont{Andreussi}},
  \bibinfo{author}{\bibfnamefont{T.}~\bibnamefont{Brumme}},
  \bibinfo{author}{\bibfnamefont{O.}~\bibnamefont{Bunau}},
  \bibinfo{author}{\bibfnamefont{M.~B.} \bibnamefont{Nardelli}},
  \bibinfo{author}{\bibfnamefont{M.}~\bibnamefont{Calandra}},
  \bibinfo{author}{\bibfnamefont{R.}~\bibnamefont{Car}},
  \bibinfo{author}{\bibfnamefont{C.}~\bibnamefont{Cavazzoni}},
  \bibinfo{author}{\bibfnamefont{D.}~\bibnamefont{Ceresoli}},
  \bibinfo{author}{\bibfnamefont{M.}~\bibnamefont{Cococcioni}},
  \bibnamefont{et~al.}, \bibinfo{journal}{J. Phys. Condens. Matter}
  \textbf{\bibinfo{volume}{29}}, \bibinfo{pages}{465901}
  (\bibinfo{year}{2017}).

\bibitem[{\citenamefont{Hamann et~al.}(1979)\citenamefont{Hamann, Schl\"uter,
  and Chiang}}]{nc}
\bibinfo{author}{\bibfnamefont{D.~R.} \bibnamefont{Hamann}},
  \bibinfo{author}{\bibfnamefont{M.}~\bibnamefont{Schl\"uter}},
  \bibnamefont{and} \bibinfo{author}{\bibfnamefont{C.}~\bibnamefont{Chiang}},
  \bibinfo{journal}{Phys. Rev. Lett.} \textbf{\bibinfo{volume}{43}},
  \bibinfo{pages}{1494} (\bibinfo{year}{1979}).

\bibitem[{\citenamefont{Perdew et~al.}(1996)\citenamefont{Perdew, Burke, and
  Ernzerhof}}]{Perdew1996}
\bibinfo{author}{\bibfnamefont{J.~P.} \bibnamefont{Perdew}},
  \bibinfo{author}{\bibfnamefont{K.}~\bibnamefont{Burke}}, \bibnamefont{and}
  \bibinfo{author}{\bibfnamefont{M.}~\bibnamefont{Ernzerhof}},
  \bibinfo{journal}{Phys. Rev. Lett.} \textbf{\bibinfo{volume}{77}},
  \bibinfo{pages}{3865} (\bibinfo{year}{1996}).

\bibitem[{\citenamefont{Togo and Tanaka}(2015)}]{phonopy}
\bibinfo{author}{\bibfnamefont{A.}~\bibnamefont{Togo}} \bibnamefont{and}
  \bibinfo{author}{\bibfnamefont{I.}~\bibnamefont{Tanaka}},
  \bibinfo{journal}{Scr. Mater.} \textbf{\bibinfo{volume}{108}},
  \bibinfo{pages}{1} (\bibinfo{year}{2015}).

\bibitem[{\citenamefont{Li et~al.}(2014)\citenamefont{Li, Carrete, Katcho, and
  Mingo}}]{ShengBTE}
\bibinfo{author}{\bibfnamefont{W.}~\bibnamefont{Li}},
  \bibinfo{author}{\bibfnamefont{J.}~\bibnamefont{Carrete}},
  \bibinfo{author}{\bibfnamefont{N.~A.} \bibnamefont{Katcho}},
  \bibnamefont{and} \bibinfo{author}{\bibfnamefont{N.}~\bibnamefont{Mingo}},
  \bibinfo{journal}{Comp. Phys. Commun.} \textbf{\bibinfo{volume}{185}},
  \bibinfo{pages}{1747} (\bibinfo{year}{2014}).

\bibitem[{\citenamefont{Li et~al.}(2012)\citenamefont{Li, Lindsay, Broido,
  Stewart, and Mingo}}]{thirdorder}
\bibinfo{author}{\bibfnamefont{W.}~\bibnamefont{Li}},
  \bibinfo{author}{\bibfnamefont{L.}~\bibnamefont{Lindsay}},
  \bibinfo{author}{\bibfnamefont{D.~A.} \bibnamefont{Broido}},
  \bibinfo{author}{\bibfnamefont{D.~A.} \bibnamefont{Stewart}},
  \bibnamefont{and} \bibinfo{author}{\bibfnamefont{N.}~\bibnamefont{Mingo}},
  \bibinfo{journal}{Phys. Rev. B} \textbf{\bibinfo{volume}{86}},
  \bibinfo{pages}{174307} (\bibinfo{year}{2012}).

\bibitem[{\citenamefont{Mostofi et~al.}(2014)\citenamefont{Mostofi, Yates,
  Pizzi, Lee, Souza, Vanderbilt, and Marzari}}]{wannier}
\bibinfo{author}{\bibfnamefont{A.~A.} \bibnamefont{Mostofi}},
  \bibinfo{author}{\bibfnamefont{J.~R.} \bibnamefont{Yates}},
  \bibinfo{author}{\bibfnamefont{G.}~\bibnamefont{Pizzi}},
  \bibinfo{author}{\bibfnamefont{Y.-S.} \bibnamefont{Lee}},
  \bibinfo{author}{\bibfnamefont{I.}~\bibnamefont{Souza}},
  \bibinfo{author}{\bibfnamefont{D.}~\bibnamefont{Vanderbilt}},
  \bibnamefont{and} \bibinfo{author}{\bibfnamefont{N.}~\bibnamefont{Marzari}},
  \bibinfo{journal}{Comput. Phys. Commun.} \textbf{\bibinfo{volume}{185}},
  \bibinfo{pages}{2309} (\bibinfo{year}{2014}).

\bibitem[{\citenamefont{Poncé et~al.}(2016)\citenamefont{Poncé, Margine,
  Verdi, and Giustino}}]{EPW}
\bibinfo{author}{\bibfnamefont{S.}~\bibnamefont{Poncé}},
  \bibinfo{author}{\bibfnamefont{E.}~\bibnamefont{Margine}},
  \bibinfo{author}{\bibfnamefont{C.}~\bibnamefont{Verdi}}, \bibnamefont{and}
  \bibinfo{author}{\bibfnamefont{F.}~\bibnamefont{Giustino}},
  \bibinfo{journal}{Comput. Phys. Commun.} \textbf{\bibinfo{volume}{209}},
  \bibinfo{pages}{116 } (\bibinfo{year}{2016}), ISSN \bibinfo{issn}{0010-4655}.

\bibitem[{\citenamefont{Wu et~al.}(2018{\natexlab{a}})\citenamefont{Wu, Zhang,
  Song, Troyer, and Soluyanov}}]{wcc}
\bibinfo{author}{\bibfnamefont{Q.}~\bibnamefont{Wu}},
  \bibinfo{author}{\bibfnamefont{S.}~\bibnamefont{Zhang}},
  \bibinfo{author}{\bibfnamefont{H.-F.} \bibnamefont{Song}},
  \bibinfo{author}{\bibfnamefont{M.}~\bibnamefont{Troyer}}, \bibnamefont{and}
  \bibinfo{author}{\bibfnamefont{A.~A.} \bibnamefont{Soluyanov}},
  \bibinfo{journal}{Comput. Phys. Commun.} \textbf{\bibinfo{volume}{224}},
  \bibinfo{pages}{405} (\bibinfo{year}{2018}{\natexlab{a}}), ISSN
  \bibinfo{issn}{0010-4655}.

\bibitem[{\citenamefont{Zheng et~al.}(2003)\citenamefont{Zheng, Payne, Feng,
  and Lim}}]{PhysRevB.67.153105}
\bibinfo{author}{\bibfnamefont{J.-C.} \bibnamefont{Zheng}},
  \bibinfo{author}{\bibfnamefont{M.~C.} \bibnamefont{Payne}},
  \bibinfo{author}{\bibfnamefont{Y.~P.} \bibnamefont{Feng}}, \bibnamefont{and}
  \bibinfo{author}{\bibfnamefont{A.~T.-L.} \bibnamefont{Lim}},
  \bibinfo{journal}{Phys. Rev. B} \textbf{\bibinfo{volume}{67}},
  \bibinfo{pages}{153105} (\bibinfo{year}{2003}).

\bibitem[{\citenamefont{Kang et~al.}(2019)\citenamefont{Kang, Park, Woo, and
  Kwon}}]{C9CP04372A}
\bibinfo{author}{\bibfnamefont{S.-H.} \bibnamefont{Kang}},
  \bibinfo{author}{\bibfnamefont{J.}~\bibnamefont{Park}},
  \bibinfo{author}{\bibfnamefont{S.}~\bibnamefont{Woo}}, \bibnamefont{and}
  \bibinfo{author}{\bibfnamefont{Y.-K.} \bibnamefont{Kwon}},
  \bibinfo{journal}{Phys. Chem. Chem. Phys.} \textbf{\bibinfo{volume}{21}},
  \bibinfo{pages}{24206} (\bibinfo{year}{2019}).

\bibitem[{\citenamefont{Kuc et~al.}(2011)\citenamefont{Kuc, Zibouche, and
  Heine}}]{PhysRevB.83.245213}
\bibinfo{author}{\bibfnamefont{A.}~\bibnamefont{Kuc}},
  \bibinfo{author}{\bibfnamefont{N.}~\bibnamefont{Zibouche}}, \bibnamefont{and}
  \bibinfo{author}{\bibfnamefont{T.}~\bibnamefont{Heine}},
  \bibinfo{journal}{Phys. Rev. B} \textbf{\bibinfo{volume}{83}},
  \bibinfo{pages}{245213} (\bibinfo{year}{2011}).

\bibitem[{\citenamefont{Zhang et~al.}(2018)\citenamefont{Zhang, Yin, Ding,
  Jiang, and Zhang}}]{Zhang2018}
\bibinfo{author}{\bibfnamefont{W.}~\bibnamefont{Zhang}},
  \bibinfo{author}{\bibfnamefont{J.}~\bibnamefont{Yin}},
  \bibinfo{author}{\bibfnamefont{Y.}~\bibnamefont{Ding}},
  \bibinfo{author}{\bibfnamefont{Y.}~\bibnamefont{Jiang}}, \bibnamefont{and}
  \bibinfo{author}{\bibfnamefont{P.}~\bibnamefont{Zhang}},
  \bibinfo{journal}{Nanoscale} \textbf{\bibinfo{volume}{10}},
  \bibinfo{pages}{16750} (\bibinfo{year}{2018}), ISSN \bibinfo{issn}{20403372}.

\bibitem[{\citenamefont{Takagi et~al.}(1994)\citenamefont{Takagi, Toriumi,
  Iwase, and Tango}}]{takagi}
\bibinfo{author}{\bibfnamefont{S.-I.} \bibnamefont{Takagi}},
  \bibinfo{author}{\bibfnamefont{A.}~\bibnamefont{Toriumi}},
  \bibinfo{author}{\bibfnamefont{M.}~\bibnamefont{Iwase}}, \bibnamefont{and}
  \bibinfo{author}{\bibfnamefont{H.}~\bibnamefont{Tango}},
  \bibinfo{journal}{IEEE Trans. Electron. Dev.} \textbf{\bibinfo{volume}{41}},
  \bibinfo{pages}{2357} (\bibinfo{year}{1994}).

\bibitem[{\citenamefont{Liao et~al.}(2015)\citenamefont{Liao, Zhou, Qiu,
  Dresselhaus, and Chen}}]{Liao2015}
\bibinfo{author}{\bibfnamefont{B.}~\bibnamefont{Liao}},
  \bibinfo{author}{\bibfnamefont{J.}~\bibnamefont{Zhou}},
  \bibinfo{author}{\bibfnamefont{B.}~\bibnamefont{Qiu}},
  \bibinfo{author}{\bibfnamefont{M.~S.} \bibnamefont{Dresselhaus}},
  \bibnamefont{and} \bibinfo{author}{\bibfnamefont{G.}~\bibnamefont{Chen}},
  \bibinfo{journal}{Phys. Rev. B} \textbf{\bibinfo{volume}{91}},
  \bibinfo{pages}{235419} (\bibinfo{year}{2015}).

\bibitem[{\citenamefont{Gaddemane et~al.}(2018)\citenamefont{Gaddemane,
  Vandenberghe, Van~de Put, Chen, Tiwari, Chen, and
  Fischetti}}]{PhysRevB.98.115416}
\bibinfo{author}{\bibfnamefont{G.}~\bibnamefont{Gaddemane}},
  \bibinfo{author}{\bibfnamefont{W.~G.} \bibnamefont{Vandenberghe}},
  \bibinfo{author}{\bibfnamefont{M.~L.} \bibnamefont{Van~de Put}},
  \bibinfo{author}{\bibfnamefont{S.}~\bibnamefont{Chen}},
  \bibinfo{author}{\bibfnamefont{S.}~\bibnamefont{Tiwari}},
  \bibinfo{author}{\bibfnamefont{E.}~\bibnamefont{Chen}}, \bibnamefont{and}
  \bibinfo{author}{\bibfnamefont{M.~V.} \bibnamefont{Fischetti}},
  \bibinfo{journal}{Phys. Rev. B} \textbf{\bibinfo{volume}{98}},
  \bibinfo{pages}{115416} (\bibinfo{year}{2018}).

\bibitem[{\citenamefont{Guo et~al.}(2019)\citenamefont{Guo, Liu, Zhu, and
  Zheng}}]{MoS2_mu}
\bibinfo{author}{\bibfnamefont{F.}~\bibnamefont{Guo}},
  \bibinfo{author}{\bibfnamefont{Z.}~\bibnamefont{Liu}},
  \bibinfo{author}{\bibfnamefont{M.}~\bibnamefont{Zhu}}, \bibnamefont{and}
  \bibinfo{author}{\bibfnamefont{Y.}~\bibnamefont{Zheng}},
  \bibinfo{journal}{Phys. Chem. Chem. Phys.} \textbf{\bibinfo{volume}{21}},
  \bibinfo{pages}{22879} (\bibinfo{year}{2019}).

\bibitem[{\citenamefont{Gu and Yang}(2015)}]{doi:10.1063/1.4905540}
\bibinfo{author}{\bibfnamefont{X.}~\bibnamefont{Gu}} \bibnamefont{and}
  \bibinfo{author}{\bibfnamefont{R.}~\bibnamefont{Yang}}, \bibinfo{journal}{J.
  Appl. Phys.} \textbf{\bibinfo{volume}{117}}, \bibinfo{pages}{025102}
  (\bibinfo{year}{2015}).

\bibitem[{\citenamefont{Cai et~al.}(2014)\citenamefont{Cai, Lan, Zhang, and
  Zhang}}]{YC1}
\bibinfo{author}{\bibfnamefont{Y.}~\bibnamefont{Cai}},
  \bibinfo{author}{\bibfnamefont{J.}~\bibnamefont{Lan}},
  \bibinfo{author}{\bibfnamefont{G.}~\bibnamefont{Zhang}}, \bibnamefont{and}
  \bibinfo{author}{\bibfnamefont{Y.-W.} \bibnamefont{Zhang}},
  \bibinfo{journal}{Phys. Rev. B} \textbf{\bibinfo{volume}{89}},
  \bibinfo{pages}{035438} (\bibinfo{year}{2014}).

\bibitem[{\citenamefont{Peng et~al.}(2016)\citenamefont{Peng, Zhang, Shao, Xu,
  Zhang, and Zhu}}]{Peng2015}
\bibinfo{author}{\bibfnamefont{B.}~\bibnamefont{Peng}},
  \bibinfo{author}{\bibfnamefont{H.}~\bibnamefont{Zhang}},
  \bibinfo{author}{\bibfnamefont{H.}~\bibnamefont{Shao}},
  \bibinfo{author}{\bibfnamefont{Y.}~\bibnamefont{Xu}},
  \bibinfo{author}{\bibfnamefont{X.}~\bibnamefont{Zhang}}, \bibnamefont{and}
  \bibinfo{author}{\bibfnamefont{H.}~\bibnamefont{Zhu}}, \bibinfo{journal}{RSC
  Adv.} \textbf{\bibinfo{volume}{6}}, \bibinfo{pages}{5767}
  (\bibinfo{year}{2016}).

\bibitem[{\citenamefont{Yan et~al.}(2014)\citenamefont{Yan, Simpson,
  Bertolazzi, Brivio, Watson, Wu, Kis, Luo, {Hight Walker}, and
  Xing}}]{Yan2014}
\bibinfo{author}{\bibfnamefont{R.}~\bibnamefont{Yan}},
  \bibinfo{author}{\bibfnamefont{J.~R.} \bibnamefont{Simpson}},
  \bibinfo{author}{\bibfnamefont{S.}~\bibnamefont{Bertolazzi}},
  \bibinfo{author}{\bibfnamefont{J.}~\bibnamefont{Brivio}},
  \bibinfo{author}{\bibfnamefont{M.}~\bibnamefont{Watson}},
  \bibinfo{author}{\bibfnamefont{X.}~\bibnamefont{Wu}},
  \bibinfo{author}{\bibfnamefont{A.}~\bibnamefont{Kis}},
  \bibinfo{author}{\bibfnamefont{T.}~\bibnamefont{Luo}},
  \bibinfo{author}{\bibfnamefont{A.~R.} \bibnamefont{{Hight Walker}}},
  \bibnamefont{and} \bibinfo{author}{\bibfnamefont{H.~G.} \bibnamefont{Xing}},
  \bibinfo{journal}{ACS Nano} \textbf{\bibinfo{volume}{8}},
  \bibinfo{pages}{986} (\bibinfo{year}{2014}).

\bibitem[{\citenamefont{Lee et~al.}(2019)\citenamefont{Lee, Kang, and
  Kwon}}]{seungjun}
\bibinfo{author}{\bibfnamefont{S.}~\bibnamefont{Lee}},
  \bibinfo{author}{\bibfnamefont{S.-H.} \bibnamefont{Kang}}, \bibnamefont{and}
  \bibinfo{author}{\bibfnamefont{Y.-K.} \bibnamefont{Kwon}},
  \bibinfo{journal}{Sci. Rep.} \textbf{\bibinfo{volume}{9}},
  \bibinfo{pages}{5149} (\bibinfo{year}{2019}).

\bibitem[{\citenamefont{Qi and Zhang}(2011)}]{RevModPhys.83.1057}
\bibinfo{author}{\bibfnamefont{X.-L.} \bibnamefont{Qi}} \bibnamefont{and}
  \bibinfo{author}{\bibfnamefont{S.-C.} \bibnamefont{Zhang}},
  \bibinfo{journal}{Rev. Mod. Phys.} \textbf{\bibinfo{volume}{83}},
  \bibinfo{pages}{1057} (\bibinfo{year}{2011}).

\bibitem[{\citenamefont{Hasan and Kane}(2010)}]{RevModPhys.82.3045}
\bibinfo{author}{\bibfnamefont{M.~Z.} \bibnamefont{Hasan}} \bibnamefont{and}
  \bibinfo{author}{\bibfnamefont{C.~L.} \bibnamefont{Kane}},
  \bibinfo{journal}{Rev. Mod. Phys.} \textbf{\bibinfo{volume}{82}},
  \bibinfo{pages}{3045} (\bibinfo{year}{2010}).

\bibitem[{\citenamefont{Wu et~al.}(2018{\natexlab{b}})\citenamefont{Wu, Zhang,
  Song, Troyer, and Soluyanov}}]{wt}
\bibinfo{author}{\bibfnamefont{Q.}~\bibnamefont{Wu}},
  \bibinfo{author}{\bibfnamefont{S.}~\bibnamefont{Zhang}},
  \bibinfo{author}{\bibfnamefont{H.-F.} \bibnamefont{Song}},
  \bibinfo{author}{\bibfnamefont{M.}~\bibnamefont{Troyer}}, \bibnamefont{and}
  \bibinfo{author}{\bibfnamefont{A.~A.} \bibnamefont{Soluyanov}},
  \bibinfo{journal}{Comput. Phys. Commun.} \textbf{\bibinfo{volume}{224}},
  \bibinfo{pages}{405 } (\bibinfo{year}{2018}{\natexlab{b}}), ISSN
  \bibinfo{issn}{0010-4655}.

\bibitem[{\citenamefont{Taghizadeh~Sisakht
  et~al.}(2016)\citenamefont{Taghizadeh~Sisakht, Fazileh, Zare, Zarenia, and
  Peeters}}]{PhysRevB.94.085417}
\bibinfo{author}{\bibfnamefont{E.}~\bibnamefont{Taghizadeh~Sisakht}},
  \bibinfo{author}{\bibfnamefont{F.}~\bibnamefont{Fazileh}},
  \bibinfo{author}{\bibfnamefont{M.~H.} \bibnamefont{Zare}},
  \bibinfo{author}{\bibfnamefont{M.}~\bibnamefont{Zarenia}}, \bibnamefont{and}
  \bibinfo{author}{\bibfnamefont{F.~M.} \bibnamefont{Peeters}},
  \bibinfo{journal}{Phys. Rev. B} \textbf{\bibinfo{volume}{94}},
  \bibinfo{pages}{085417} (\bibinfo{year}{2016}).

\bibitem[{\citenamefont{Liu et~al.}(2015)\citenamefont{Liu, Zhang, Abdalla,
  Fazzio, and Zunger}}]{doi:10.1021/nl5043769}
\bibinfo{author}{\bibfnamefont{Q.}~\bibnamefont{Liu}},
  \bibinfo{author}{\bibfnamefont{X.}~\bibnamefont{Zhang}},
  \bibinfo{author}{\bibfnamefont{L.~B.} \bibnamefont{Abdalla}},
  \bibinfo{author}{\bibfnamefont{A.}~\bibnamefont{Fazzio}}, \bibnamefont{and}
  \bibinfo{author}{\bibfnamefont{A.}~\bibnamefont{Zunger}},
  \bibinfo{journal}{Nano Lett.} \textbf{\bibinfo{volume}{15}},
  \bibinfo{pages}{1222} (\bibinfo{year}{2015}).

\end{thebibliography}


\begin{thebibliography}{0}%
\makeatletter
\providecommand \@ifxundefined [1]{%
 \@ifx{#1\undefined}
}%
\providecommand \@ifnum [1]{%
 \ifnum #1\expandafter \@firstoftwo
 \else \expandafter \@secondoftwo
 \fi
}%
\providecommand \@ifx [1]{%
 \ifx #1\expandafter \@firstoftwo
 \else \expandafter \@secondoftwo
 \fi
}%
\providecommand \natexlab [1]{#1}%
\providecommand \enquote  [1]{``#1''}%
\providecommand \bibnamefont  [1]{#1}%
\providecommand \bibfnamefont [1]{#1}%
\providecommand \citenamefont [1]{#1}%
\providecommand \href@noop [0]{\@secondoftwo}%
\providecommand \href [0]{\begingroup \@sanitize@url \@href}%
\providecommand \@href[1]{\@@startlink{#1}\@@href}%
\providecommand \@@href[1]{\endgroup#1\@@endlink}%
\providecommand \@sanitize@url [0]{\catcode `\\12\catcode `\$12\catcode
  `\&12\catcode `\#12\catcode `\^12\catcode `\_12\catcode `\%12\relax}%
\providecommand \@@startlink[1]{}%
\providecommand \@@endlink[0]{}%
\providecommand \url  [0]{\begingroup\@sanitize@url \@url }%
\providecommand \@url [1]{\endgroup\@href {#1}{\urlprefix }}%
\providecommand \urlprefix  [0]{URL }%
\providecommand \Eprint [0]{\href }%
\providecommand \doibase [0]{http://dx.doi.org/}%
\providecommand \selectlanguage [0]{\@gobble}%
\providecommand \bibinfo  [0]{\@secondoftwo}%
\providecommand \bibfield  [0]{\@secondoftwo}%
\providecommand \translation [1]{[#1]}%
\providecommand \BibitemOpen [0]{}%
\providecommand \bibitemStop [0]{}%
\providecommand \bibitemNoStop [0]{.\EOS\space}%
\providecommand \EOS [0]{\spacefactor3000\relax}%
\providecommand \BibitemShut  [1]{\csname bibitem#1\endcsname}%
\let\auto@bib@innerbib\@empty
%</preamble>
\end{thebibliography}%
\end{document}

%\listoftodos
%\newpage

Supplementary Information \\

\title{Versatile Physical Properties of a Novel Two-Dimensional Materials Composed of Group IV-V Elements}

\author{Seungjun Lee}
\affiliation{Department of Physics and
             Research Institute for Basic Sciences,
             Kyung Hee University, Seoul, 02447, Korea}

\author{Young-Kyun Kwon}
\email[Corresponding author. E-mail: ]{ykkwon@khu.ac.kr}
\affiliation{Department of Physics and
             Research Institute for Basic Sciences,
             Kyung Hee University, Seoul, 02447, Korea}

\date{\today}
\maketitle
%---------------------------------------------------------------------
%---------------------------------------------------------------------
\begin{center}
\begin{table}
\caption {Lattice constant (\AA) of mirror and inversion phases of group IV-V  compounds with primitive 2D hexagonal unit cells
\label{table1}}
\begin{ruledtabular}
\begin{tabular}{ c c c c c c c c c } 
  \multirow{2}{3em}{} & \multicolumn{4}{c}{Mirror} & \multicolumn{4}{c}{Inversion}\\
  & C & Si & Ge & Sn & C & Si & Ge & Sn \\
    \hline
N  &2.39 &2.90 &3.16 &3.42 &2.40 &2.91 &3.18 &3.43 \\
P  &2.89 &3.52 &3.69 &3.94 &2.90 &3.54 &3.70 &3.95 \\
As &3.14 &3.90 &3.85 &4.90 &3.15 &3.71 &3.86 &4.10 \\
Sn &3.40 &4.00 &4.13 &4.37 &3.41 &4.01 &4.14 &4.38 \\
Bi &3.59 &4.17 &4.27 &4.52 &3.60 &4.18 &4.28 &4.53 \\
\end{tabular}
\end{ruledtabular}
\end{table}
\end{center}
%---------------------------------------------------------------------
%---------------------------------------------------------------------
\begin{center}
\begin{table}
\caption {Cohesive energies (eV/atom) of all compounds. Each value was calculated using the total energy differences between the group IV-V compounds and isolated atoms.
\label{table2}}
\begin{ruledtabular}
\begin{tabular}{ c c c c c c c c c } 
  \multirow{2}{3em}{} & \multicolumn{4}{c}{Mirror} & \multicolumn{4}{c}{Inversion}\\
  & C & Si & Ge & Sn & C & Si & Ge & Sn \\
    \hline
N  &-5.84 &-5.44 &-3.81 &-3.61 &-5.85 &-5.43 &-3.81 &-3.60\\
P  &-5.25 &-4.19 &-3.48 &-3.27 &-5.27 &-4.19 &-3.47 &-3.26\\
As &-4.60 &-3.81 &-3.24 &-3.07 &-4.63 &-3.80 &-3.22 &-3.06\\
Sn &-4.34 &-3.54 &-3.06 &-2.90 &-4.38 &-3.52 &-3.05 &-2.89\\
Bi &-4.06 &-3.31 &-3.03 &-2.79 &-4.11 &-3.31 &-3.01 &-2.77\\
\end{tabular}
\end{ruledtabular}
\end{table}
\end{center}
% entire page. There is no need to do explicit centering.
\begin{figure}[t]
\includegraphics[width=1.0\columnwidth]{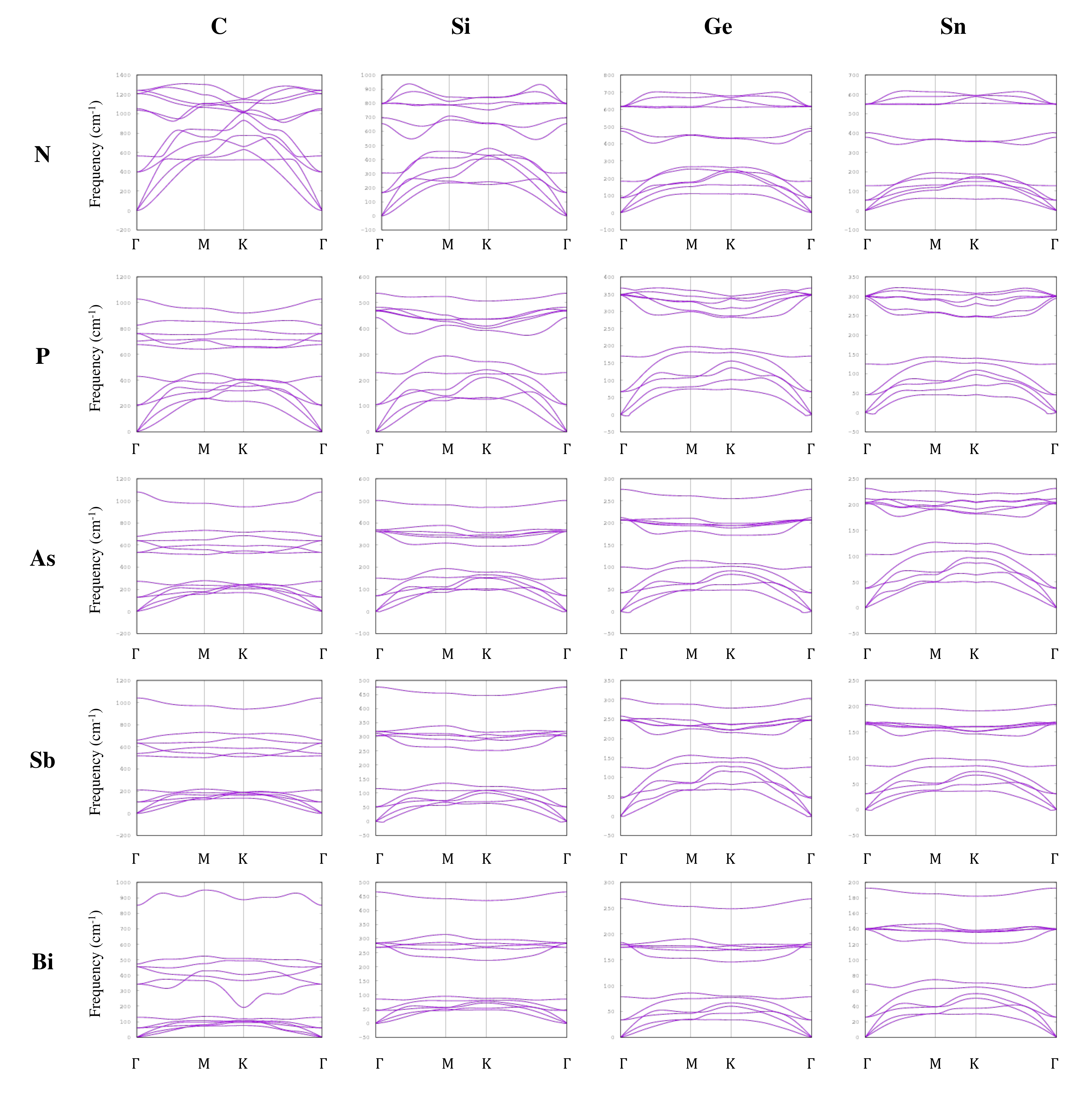}
\caption{Calculated phonon dispersion relations of $\mathcal{M}$-A$_2$B$_2$
\label{S_phonon1}}
\end{figure}
% entire page. There is no need to do explicit centering.
\begin{figure}[t]
\includegraphics[width=1.0\columnwidth]{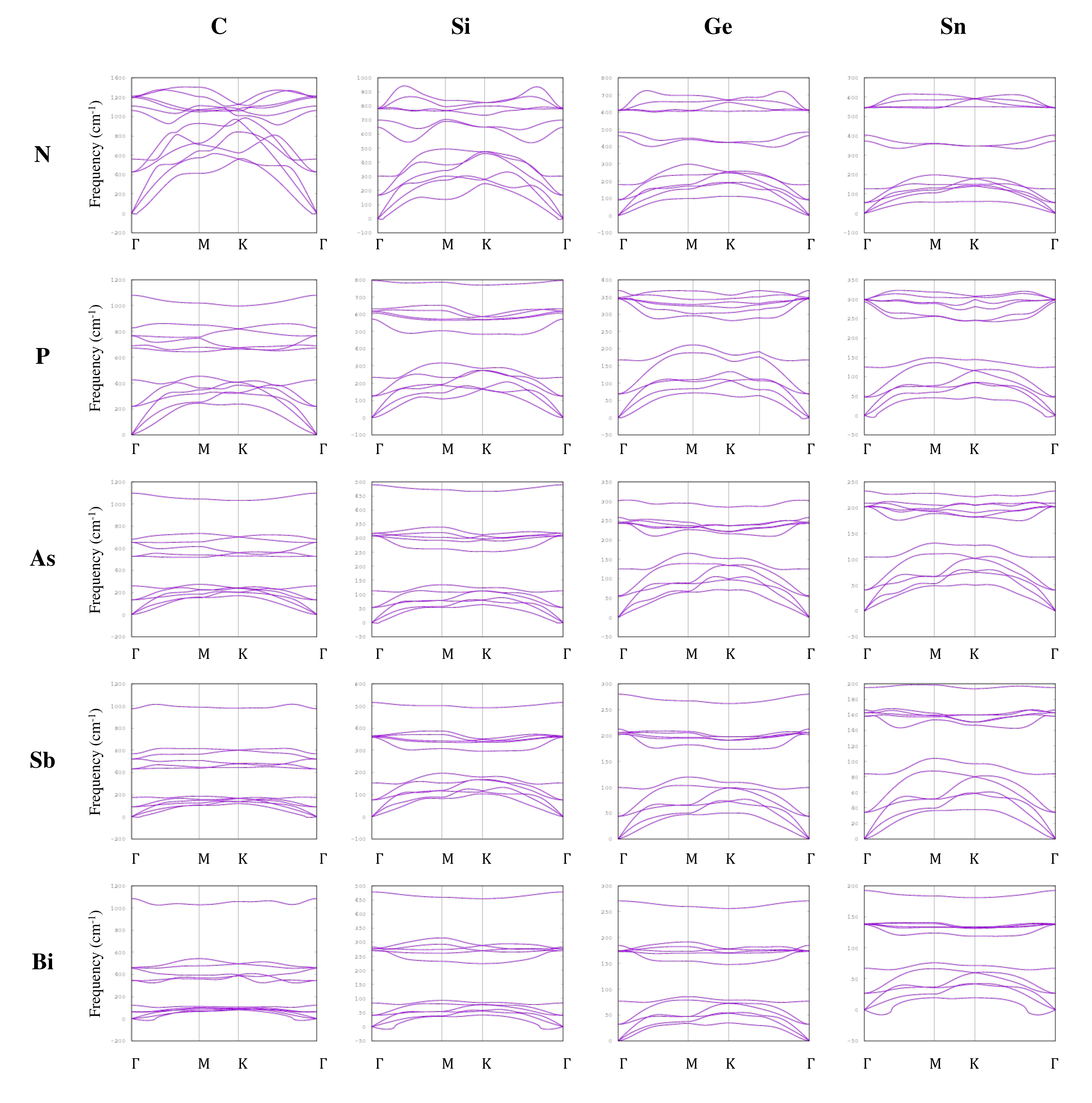}
\caption{Calculated phonon dispersion relations of $\mathcal{I}$-A$_2$B$_2$
\label{S_phonon2}}
\end{figure}
% entire page. There is no need to do explicit centering.
\begin{figure}[t]
\includegraphics[width=1.0\columnwidth]{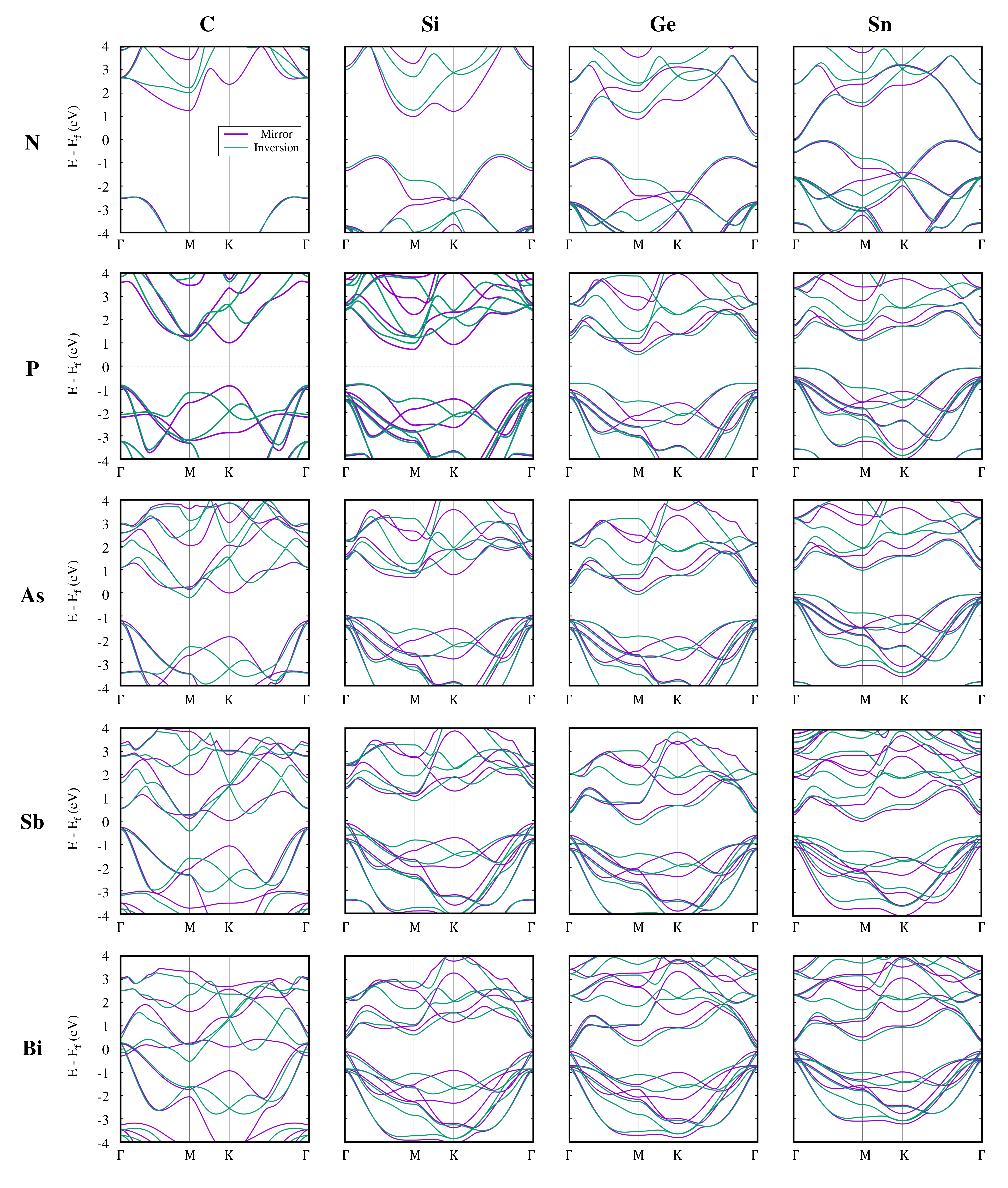}
\caption{Calculated band structures of $\mathcal{M}$- and $\mathcal{I}$ -A$_2$B$_2$. The purple and green lines indicate $\mathcal{M}$ and $\mathcal{I}$ phases, respectively.
\label{S_band}}
\end{figure}
% entire page. There is no need to do explicit centering.
\begin{figure}[t]
\includegraphics[width=1.0\columnwidth]{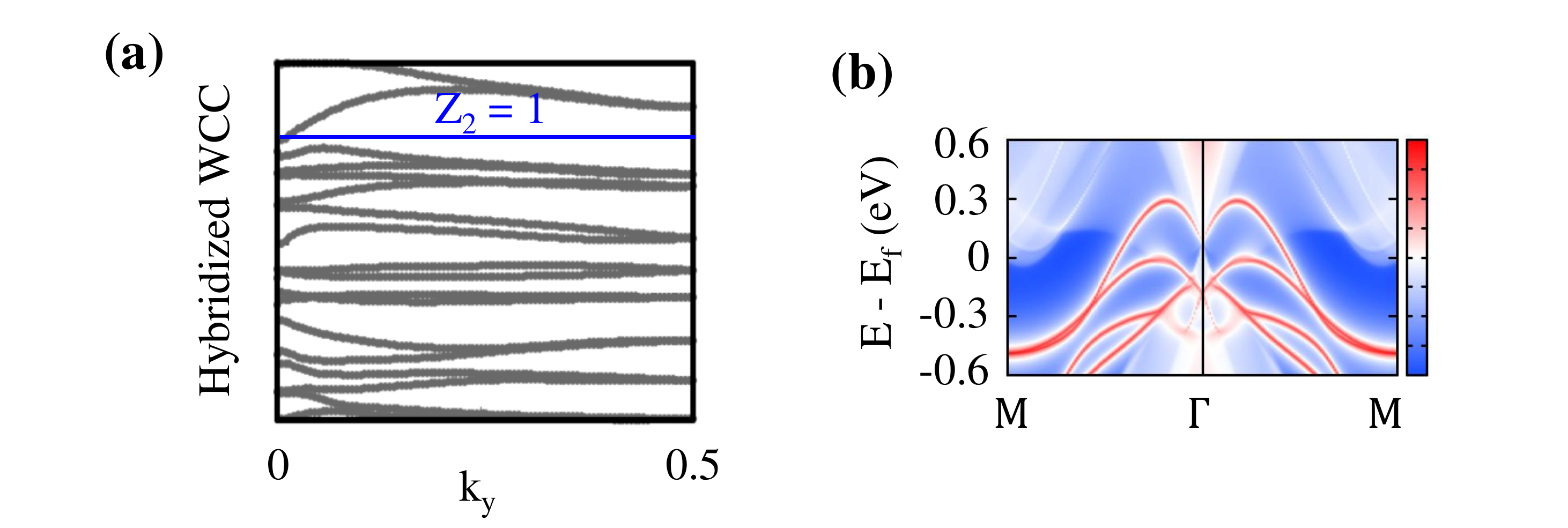}
\caption{(a) Evolution of the hybridized Wannier charge center,
and (b) surface band structure of $\mathcal{M}$-Sn$_2$Bi$_2$. 
\label{TI_M}}
\end{figure}
%---------------------------------------------------------------------
%%---------------------------------------------------------------------
%\begin{center}
%\begin{table}
%\caption {Room-temperature carrier mobilities of Si$_2$P$_2$ and C$_2$P$_2$ for the both $\mathcal{M}$ and $\mathcal{I}$ phases at the moderate carrier concentration (10$^{12}$cm$^{-2}$)
%\label{table_mu}}
%\begin{ruledtabular}
%\begin{tabular}{ccccc} 
%  \multirow{2}{3em}{} & \multicolumn{2}{c}{Si$_2$P$_2$} & \multicolumn{2}{c}{C$_2$P$_2$}\\
%  & $\mathcal{M}$ & $\mathcal{I}$ & $\mathcal{M}$ & $\mathcal{I}$ \\
%    \hline
%electron  & ?? & ?? & ?? & ?? \\
%hole      & ?? & ?? & ?? & ?? \\
%\end{tabular}
%\end{ruledtabular}
%\end{table}
%\end{center}
%%---------------------------------------------------------------------
%We also summarized their room-temperature carrier mobilities in the  Table~\ref{table_mu}. 